\documentclass[11pt,a4paper]{article}
\usepackage{jheppub}

\usepackage{bbm}
\def\beq{\begin{equation}}
\def\eeq{\end{equation}}
\def\bea{\begin{eqnarray}}
\def\eea{\end{eqnarray}}

\def\tr{\textrm{tr}}

\def\d{\partial}
\def\cM{\mathcal{M}}

\def\cT{\mathcal{T}}

\def\cO{\mathcal{O}}

\def\bb{\bar{b}}

\def\dt{\partial}

\title{ Completing the framework of AdS/QCD: $h_1/b_1$ mesons and excited $\omega/\rho$'s}

\author[a]{S.~K.~Domokos,}
\author[b]{J.~A.~Harvey,}
\author[c]{A.~B.~Royston}

\affiliation[a]{Department of Particle Physics and Astrophysics\\ Weizmann Institute of Science, Rehovot 76100, Israel}
\affiliation[b]{Enrico Fermi Institute and Department of Physics \\5640 Ellis Ave., Chicago IL 60637, USA}
\affiliation[c]{NHETC and Department of Physics and Astronomy, Rutgers University \\ 126 Frelinghuysen Rd., Piscataway NJ 08855, USA}

\emailAdd{sophia.domokos@weizmann.ac.il}
\emailAdd{j-harvey@uchicago.edu}
\emailAdd{aroyston@physics.rutgers.edu}

\abstract{We extend the ``hard wall'' gravity dual of QCD by
including tensor fields $b_{MN}$ that correspond to the QCD quark
bilinear operators $\bar q \sigma^{\mu \nu} q$. These fields
give rise to a spectrum of states which include the $h_1$ and $b_1$
mesons, as well as a tower of excited $\omega/\rho$ meson states. We
also identify the lowest-dimension term which leads to mixing
between the new $\rho$ states and the usual tower of $\rho$ mesons
when chiral symmetry is broken. }

\keywords{AdS-CFT Correspondence, AdS/QCD}

\begin{document}
\begin{flushright} WIS/01/11-JAN-DPPA \\ EFI-11-2 \\ RUNHETC-2010-26 \end{flushright}

\maketitle

\section{Introduction and motivation}

The last few years have seen a renewed focus on string dual
descriptions of QCD following the discovery of the AdS/CFT
correspondence  \cite{Maldacena,AdSCFT,AdSCFT2} and subsequent
attempts to construct holographic string/gravity duals of QCD in
both top-down \cite{SakaiI,SakaiII} and bottom-up approaches
\cite{Son:2003et,hardwall,DaRoldI,DaRoldII,HirnSanz}. These models
have been remarkably successful in reproducing some of the
low-energy features of QCD, but also suffer a number of flaws
discussed at length in the literature
\cite{Shifman,GlozmanI,GlozmanII,Csaki}.

For concreteness, let us consider the ``hard wall'' model of
\cite{hardwall}. This model contains five-dimensional (5d) fields $A_{L,\mu}^a$,
$A_{R,\mu}^a$ and $X^{\alpha \beta}$ dual to the operators  $\bar
q_L \gamma_\mu t^a q_L$, $\bar q_R \gamma_\mu t^a q_R$ and $\bar
q_R^\alpha q_L^\beta$ respectively. These fields live in a 5d Anti
de Sitter ($AdS$) space with a hard cutoff on the radial coordinate at
$z=z_0$ with $1/z_0 = 346 ~ {\rm MeV}$. The model gives rise to a
spectrum of 4d fields whose overall scale is determined by the
infrared (IR) cutoff $z_0$. The lightest modes of definite parity
are the $\pi$, $\rho$ and $a_1$ mesons; the model generates an
excellent fit to the masses, decay constants and coupling constants
of these mesons in terms of three parameters: $z_0$, the quark mass
$m_q=2.3 ~{\rm MeV}$, and the scale of chiral symmetry breaking
$\sigma= (308~ {\rm MeV})^3$ (using the Model B values of
\cite{hardwall}). Other ``bottom-up'' models
\cite{DaRoldI,DaRoldII,HirnSanz} have a very similar structure. The
``top-down'' model of \cite{SakaiI} is based on a D-brane
construction in string theory, and whose field content is
uniquely as a result. It is roughly as successful at predicting the properties
of the low-lying mesons as the bottom-up approaches.

Despite its successes, the hard wall model suffers from a variety of issues, some of which we address
in this paper.
 The most obvious problem has to do with the structure of the excited states in the meson spectrum.
 The first excited $\rho$  or $\omega$ meson
state is predicted to lie at $m(\rho')=x_{0,2}/z_0= 1910 ~{\rm MeV}$
where $x_{0,2}$ is the second zero of the Bessel function $J_0(x)$,
while in reality the mass of the observed excited state is at
roughly $1450 ~{\rm MeV}$. In addition, the mass squared of higher
excited states scales with the excitation number as $n^2$ rather
than as $n$ as expected from Regge theory and semi-classical
arguments \cite{Shifman,softwall}.

One solution to the problem of the excited meson spectrum, analyzed
in \cite{softwall}, is to alter the $z$-dependence of the metric and
of the dilaton profile. In \cite{softwall} a simple modification was
suggested, which leads to a spectrum of excited vector mesons of the
form $m_n^2= \Sigma(n+1)$ where $\Sigma$ is a constant related to
the QCD string tension. For an appropriate choice of $\Sigma$ this
fits the low-lying spectrum reasonably well and leads to the
expected asymptotic behavior at large $n$.

However, we will see that this cannot be the full story. The hard
wall model does not even fully describe the light meson spectrum. It
gives reasonable predictions for the lowest-lying isotriplet vector
and axial-vector mesons, the $\rho$ and $a_1$ (and also the
isosinglet $\omega$ and $f_1$ if the gauge group is generalized from
$SU(2)_L \times  SU(2)_R$ to $U(2)_L \times U(2)_R$)  but does not
include fields which give rise to the $J^{PC}=1^{+-}$ $h_1$ and
$b_1$ mesons whose masses, at $ \sim 1200 ~{\rm MeV}$, lie just
below those of the $a_1$ and $f_1$. Furthermore, as discussed in
\cite{Chizhov,Shifman,GlozmanI}, from QCD one actually expects two types of
$\rho$ mesons. The first type couples dominantly to the usual vector
current $\bar q \gamma^\mu t^a q$, the second to the tensor operator
$\bar q \sigma^{\mu \nu}t^a q$. In the limit of unbroken chiral
symmetry these two operators lie in different representations of the
chiral symmetry group, leading to two distinct towers of states with
distinct (chiral) quantum numbers. When chiral symmetry is broken,
the operators mix and the physical $\rho$ mesons will be linear
combinations of the two $\rho$ varieties. One cannot expect to get a
correct spectrum of excited $\rho$ and $\omega$ mesons if this
structure is ignored.

A possible solution for including the $h_1$ and $b_1$ mesons was
suggested in \cite{Erlich}; one should add 5d fields dual to the
interpolating operators for these mesons.  These operators are of
the form $\bar q \sigma^{\mu \nu} t^a q$; incorporating their dual
fields is a very natural generalization of the hard wall and other
dual models of QCD. Once one includes fields dual to some of the
canonical dimension three operators in QCD as was done in
\cite{hardwall} it is natural to include fields dual to {\it all}
the canonical dimension three operators in QCD. Although the
solution is conceptually clear, we will need to overcome a number of
technical issues in order to implement this idea.

In what follows we will argue that there is a natural extension of
the hard wall model which includes complex, antisymmetric tensor
fields $b_{MN}$ transforming in the bifundamental of the $U(N_f)_L
\times U(N_f)_R$ gauge symmetry, that are dual to the operators
$\bar q \sigma^{\mu \nu} t^a q$\footnote{We work in the context of
the hard wall model for simplicity. A more complete dual description
of QCD will undoubtedly incorporate the tensor fields discussed here
as well as modifications of the metric and dilaton profile in order
to produce the correct asymptotic behavior of excited states.  The
dual description of the thermodynamic properties of QCD also
requires such modifications. See e.g. \cite{kiritsis} for a recent
review.}.  We will show that these fields have a natural first order
action, familiar from the description of charged tensor fields in
$AdS_5 \times S^5$ \cite{AFII,Minces},
and that their expansion in terms of 4d fields leads to a tower of
mesons with the quantum numbers of the $h_1$ and $b_1$ as well as an
additional tower of mesons with the same quantum numbers as the
$\rho$ and $\omega$ mesons.
We will also identify the lowest dimension operator which induces
mixing between the two distinct towers of $\rho /\omega$ mesons in
the presence of chiral symmetry breaking, and which is also
responsible for decay processes like $h_1 \rightarrow \rho + \pi$
and $b_1 \rightarrow \omega + \pi$.  Our treatment follows the
standard formalism of string/gauge duality. In this we differ from
\cite{Cappiello} where it is claimed that modifications to the
formalism involving the asymptotic behavior of fields is required in
order to incorporate tensor mesons in the presence of chiral
symmetry breaking. Our description of tensor mesons also seems to be
somewhat at odds with the description given in \cite{Imoto}. It is
not clear to us whether the formalism used in \cite{Imoto} included
the possibility of fields with a first order Lagrangian.
\cite{softwall} does not address this additional tensor operator,
and in fact we should note that despite its encouraging fit with the
excited $\rho$ spectrum, \cite{softwall} does not address the $h_1$
and $b_1$ states. Once we include fields dual to these excitations
we necessarily generate additional vector meson states, presumably
ruining the agreement between the excited $\rho$'s of \cite{softwall}
 and experiment.

In addition to these specific issues, which are directly addressed in the body
of the paper, our analysis will also touch on some important
problems of principle that arise in the construction of QCD duals.
These involve the nature the $1/N_c$ expansion,  the validity of a
local 5d field theory description and the field theory
interpretation of the asymptotically $AdS$ spacetime.

Let us consider the last point first. It is often said that the
conformal invariance of Anti de Sitter space matches the conformal
invariance that QCD enjoys in the ultraviolet (UV) because of its
asymptotic freedom. This is not entirely correct. In the
best-understood example, the AdS/CFT correspondence, the conformal
isometry of $AdS$ space is in accordance with the conformal invariance
of $N=4$ SYM at large 't Hooft coupling. It is important to note
that in this limit $N=4$ SYM is \emph{not} a free field theory, but
is rather a nontrivial CFT with dimensions and correlators which
differ from those of free field theory. In the analysis of
\cite{hardwall} the 5d gauge coupling is determined by matching onto
the UV behavior of the vector current two point function in QCD,
which can be computed reliably from a one-loop diagram because of
asymptotic freedom. This is a very special case, however, because
the anomalous dimension of the (conserved) vector current vanishes.
In contrast, the tensor operator $\bar q t^a \sigma_{\mu \nu} q$ is
not conserved and has a nonzero anomalous dimension. We will see for
example that matching its anomalous dimension as derived from the
AdS/CFT correspondence to the UV behavior of QCD leads to an
unphysical value of the mass of the lowest $b_1/h_1$ mesons. This
suggests that the correct interpretation of models like
\cite{hardwall} is that they are modeling QCD as a theory which has
a ``conformal window" of energy scales in which the theory behaves
as a nontrivial CFT and then has conformal invariance broken at an
IR scale $1/z_0$.  This is clearly not the behavior of real world
QCD, but apparently for some purposes it is a reasonable
approximation. With this picture in mind we can obtain the
appropriate 5d mass of the tensor field (dual to the anomalous
dimension of the tensor operator) required to fit the observed mass
of the $b_1/h_1$ mesons.

We turn now to the issue of a local 5d dual description of QCD. In
current models one studies a 5d action for massless fields such as
the gauge fields $A_{L,R}$. In the top-down model of \cite{SakaiI}
these fields arise as the lightest open string modes. One
can also include the lightest closed string modes: the graviton,
dilaton and antisymmetric tensor fields, which provide a dual
description of the glueball spectrum of QCD.  One then either writes
down the lowest dimension interactions involving these fields, or
derives the lowest dimension terms using an $\alpha'$ expansion in
string theory. In other words, one works in a ``supergravity'' limit
of the full string dual. Such an approximation is usually justified
only if there is a clear separation of scales between the massive
string modes and the energy scale being used to probe the massless
string modes. There is no evidence for such a separation of scales
in real world QCD. For example, the string scale extracted from the
Regge trajectories of meson states is $\alpha' \simeq 0.88 ~{\rm
GeV}^2$ which leads us to expect  massive string modes at a scale of
order $1/\sqrt{\alpha' } \simeq 1 ~ {\rm GeV}$. Yet in both the
top-down and bottom-up models this scale is comparable to the scale
of excited states of the massless modes. Put another way, in the
current models one keeps only the lowest dimension operators
constructed out of the massless modes. If there is no separation of
energy scales, there is no reason to expect higher dimension
operators to be suppressed, or for that matter, no reason to expect
that we can use a local 5d field theory description at all.

Nevertheless, in what follows we assume that it makes sense to use a
local 5d field theory description. One possible theoretical
justification for this has been suggested in
\cite{wittentalk,softwall}. \cite{wittentalk} proposes that there
exists a local 5d description in the $N \rightarrow \infty$ limit
for small 't Hooft coupling, $\lambda= g^2 N$, even though this is
the $\alpha' \rightarrow \infty$ limit rather than the $\alpha'
\rightarrow 0$ limit one usually considers to obtain a local 5d
action, and that, furthermore, the existence of such a local
description is linked to the existence of an infinite family of
conserved tensors of arbitrarily high spin. See \cite{higher} for a
review.

In \cite{softwall} it was argued that one can add 5d fields dual to
an infinite family of twist two operators in QCD of the form $\bar q
\gamma^{( \mu_1} D^{\mu_2} \cdots D^{\mu_n )}q $ while maintaining
general coordinate and tensor gauge invariance at quadratic order in
the fields. The tensor field $b_{MN}$ that we add is also dual to a
twist two operator in QCD, and we will show that again one can
maintain general coordinate and gauge invariance at quadratic order.
In fact the full leading Regge trajectory of QCD including daughter
states seems to be reasonably well-described by fields which are
dual to the twist two operators of QCD. It is also known from
semi-classical arguments that operators of fixed twist receive
corrections to their anomalous dimensions which are smaller than one
would naively expect, growing only as the logarithm of the spin $S$
\cite{gkpII}. Finally, we remind the reader of one other example
where one finds remarkably accurate results without the presence of
an obvious low-energy limit, namely level truncation in open string
field theory \cite{sen}.

We will have more to say about higher dimension terms in the
effective action when we construct the effective action for tensor
fields in section 3. Whether or not these theoretical speculations
are borne out, our fundamental justification for using a local 5d
field theory approach is pragmatic. The best way to test any model
is to extend it into new regimes, perform new calculations, and
compare it with data.

The outline of this paper is as follows. In the second section we
review basic facts about meson masses and decay constants in QCD as
well as the quantum numbers of various quark bilinear operators. In
the third section we introduce the fields dual to the dimension
three tensor operators, and construct the free 5d Lagrangian. We
then compute the spectrum and decay constants in the free theory. In
section 4 we discuss interaction terms. There are a number of
possible Lagrangians that one could consider and the proper choice
requires a careful discussion of discrete symmetries. We also
discuss the role of the $1/N_c$ expansion and dimensional analysis.
In the final section we conclude and discuss some open issues. The
appendices contain some useful technical details. In a subsequent
paper we will present a detailed analysis of chiral symmetry
breaking in this model, and the effect of the new interaction terms
on the spectrum.

\section{The QCD picture}

We begin by describing the features of  QCD that are most relevant to our analysis.

\subsection{Operators and two-point functions}

In QCD with two flavors the quark bilinear operators with naive
dimension three are
\bea \label{dim3ops}
O^{S,a}(x) &=& \bar q(x) t^a q(x)~, \cr
O^{P,a}(x) &=& \bar q(x) t^a \gamma_5 q(x)~, \cr
O_{\mu}^{V,a}(x) &=& \bar q(x) t^a \gamma_\mu q(x)~, \cr
O_{\mu}^{A,a}(x) &=& \bar q(x) t^a \gamma_\mu \gamma_5 q(x)~, \cr
O_{\mu \nu}^{T,a}(x) &=& \bar q(x) t^a \sigma_{\mu \nu} q(x)~,
\eea
with $\sigma_{\mu \nu}= i (\gamma_\mu \gamma_\nu - \gamma_\nu
\gamma_\mu)/2$. The index $a=1,2,3$ for $SU(2)$ flavor symmetry or
$a=0,1,2,3$ for $U(2)$. In what follows we mostly suppress the
flavor indices unless they are required for clarity. As the hard
wall model of \cite{hardwall} contains fields dual to $O^S, O^P,
O^V, O^A$, we emphasize here the role of the tensor operator $O^T$.

The operator $O^T$ is odd under $C$ but contains both parity even and odd parts and thus
has a non-zero amplitude to create both $J^{PC}=1^{--}$ and $1^{+-}$ states $\rho^{(n)}, b_1^{(n)}$:
\beq \label{Oodd}
\langle 0| O^{T,a}_{\mu \nu}(x) | b_1^{(n),c}(k) \rangle = \frac{i}{\sqrt{2}} f_b^{(n)} \epsilon_{\mu \nu \alpha \beta} \varepsilon_{(b)}^{(n)\alpha} k^\beta \delta^{ac} e^{-ik\cdot x}~,
\eeq
\beq \label{Oeven}
\langle 0| O^{T,a}_{\mu \nu}(x) | \rho^{(n),c}(k) \rangle = \frac{i}{\sqrt{2}} f_{\rho}^{T,(n)} (\varepsilon_{(\rho) \mu}^{(n)} k_\nu - \varepsilon_{(\rho) \nu}^{(n)} k_\mu ) \delta^{ac} e^{-ik\cdot x}~.
\eeq
where $\varepsilon_{(\rho)}^{(n)},\varepsilon_{(b)}^{(n)}$ denote transverse
polarizations. In the rest frame of the mesons with $k =(m
,\mathbf{0})$, \eqref{Oodd} implies that the $b_1$ mesons are created
by the transverse components $O^T_{ij}$ while \eqref{Oeven} implies
that the $\rho$ mesons are created by the longitudinal components
$O^T_{0i}$. This is consistent with the parity assignments of the
$b_1, \rho$.  Of course, $\rho$ states are also created by the
vector current\footnote{The literature contains many different
conventions for the normalization of decay constants, some of which
differ from those used here by factors of $2$, $\sqrt{2}$ and powers
of the meson mass. Our normalization for the decay constant $f_\rho$
is fairly standard, our normalization for the decay constants $f_b$
and $f_\rho^T$ agrees with that used in \cite{lattb1two,latticeb1}
once the difference in isospin conventions is taken into account.}
\beq \label{OVrho}
\langle 0 | O_{\mu}^{V,a}(x) | \rho^{(n),c}(k) \rangle = i m_{\rho}^{(n)} f_{\rho}^{V,(n)} \varepsilon_{(\rho) \mu}^{(n)} \delta^{ac} e^{-ik\cdot x}~.
\eeq
We will see below that there are a number of ways to split up the
degrees of freedom in $O^T$, and that different decompositions are
practical in different contexts. For instance, we can project out
the transverse and longitudinal parts of $O^T_{\mu \nu}$ in momentum
space using the transverse and longitudinal projection
operators $\mathcal{P}^ \perp$, $\mathcal{P}^\parallel$ (see Appendix \ref{ap:proj} for details) --
these isolate the $h_1/b_1$ and $\omega/\rho$-type states. We define $O^{T\perp}$ and $O^{T\parallel}$ via
$(\mathcal{P}^\perp)_{\mu\nu}^{\alpha\beta}
O^T_{\alpha \beta} = k^2 O^{T\perp}_{\mu \nu}$ and $(\mathcal{P}^\parallel)_{\mu\nu}^{\alpha\beta}
O^T_{\alpha \beta} = k^2 O^{T\parallel}_{\mu \nu}$.

On the other hand, when we discuss the transformation properties  of
$O^T$ under the chiral symmetry group $U(N_f)_L \times U(N_f)_R$ it
is convenient to decompose the tensor into self-dual and
anti-self-dual parts
\beq\label{Oasd}
O^{T,\pm}_{\mu \nu} = \bar q \sigma_{\mu \nu} \frac{1 \pm \gamma_5}{2} q~,
\eeq
which obey
\beq \label{Oop}
O^{T,\pm}_{\mu \nu} = \pm \frac{i}{2} {\epsilon_{\mu \nu}}^{\lambda \rho} O^{T,\pm}_{\lambda \rho}~.
\eeq
The operators $O^{T ,\pm}$ transform as $(1,0)$ and $(0,1)$ under Lorentz transformations and obey $(O^{T,+})^* = O^{T,-}$. That is, complex conjugation flips the spacetime chirality, as is familiar from the behavior of Weyl spinors.  Under the chiral symmetry group $O^{T,+}$ transforms as a bifundamental,
$O^{T,+} \sim ({\bf \overline{N}}_f, {\bf N}_f)$ while $O^{T,-}$ transforms as the conjugate, $({\bf N}_f, {\bf \overline{N}}_f)$.

We can also construct combinations of definite parity as
\beq \label{Otensor}
O^T_{\mu \nu}= O^+_{\mu \nu}+ O^-_{\mu \nu}~,
\eeq
which transforms as a tensor under parity and
\beq \label{Opseudotensor}
O^{PT}= O^+_{\mu \nu} - O^-_{\mu \nu}~,
\eeq
which transforms as a pseudotensor. These are not independent as a consequence of the identity
$\sigma _{\mu \nu} \gamma_5 =i  \epsilon_{\mu \nu \lambda \rho} \sigma^{\lambda \rho}$ which implies
that
\beq \label{OTPTdual}
O^{PT}_{\mu \nu} = \frac{i}{2} {\epsilon_{\mu \nu}}^{\lambda \rho} O^{T}_{\lambda \rho}~.
\eeq

Our primary source for information about the properties of the dual
theory will be the two-point function, whose momentum-space poles
mark the masses of excitations, while the residues at the poles
provide the decay constants. The large-$Q$ behavior of holographic
correlators is often compared directly to QCD. We thus summarize the
relevant QCD two-point functions here:
\bea \label{MECorRelation}
\Pi^{V,V,ab}_{\mu \nu}(k) &=& -i \int d^4x ~e^{i k \cdot x} \langle 0 | O^{V,a}_\mu(x) O^{V,b}_\nu(0) | 0 \rangle~, \cr \cr
\Pi^{T,V,ab}_{\lambda \rho,\mu}(k) &=& -i \int d^4x ~e^{i k \cdot x} \langle 0 |  O^{T,a}_{\lambda \rho}(x) O^{V,b}_\mu(0) | 0 \rangle~, \cr \cr
\Pi^{T,T,ab}_{\mu \nu, \lambda \rho}(k) & =& - i \int d^4x ~e^{i k \cdot x} \langle 0 | O^{T,a}_{\mu \nu}(x) O^{T,b}_{\lambda \rho}(0) | 0 \rangle ~.
\eea
It is conventional to separate out kinematical and group theory factors which are dictated by parity, current conservation and Lorentz invariance, so we define
\bea \label{ReducedME}
\Pi^{V,V,ab}_{\mu \nu}(k) &=& (k^2 \eta_{\mu \nu} -k_\mu k_\nu) \delta^{ab} \Pi^{V,V}(k^2)~, \cr \cr
\Pi^{T,V,ab}_{\lambda \rho,\mu}(k) &=& i(k_\lambda \eta_{\rho\mu} -k_\rho \eta_{\lambda\mu}) \delta^{ab} \Pi^{T,V}(k^2)~, \cr \cr
\Pi^{T,T,ab}_{\mu \nu, \lambda \rho}(k) &=& \delta^{ab} \left( \mathcal{P}^{\parallel}_{\mu \nu, \lambda \rho} \Pi^{T,T\parallel}(k^2) + \mathcal{P}^{\perp}_{\mu \nu, \lambda \rho} \Pi^{T,T\perp}(k^2) \right)~.
\eea

These two point functions receive both perturbative and nonperturbative contributions and have been studied in \cite{svz,Reinders,Govaerts,CataI,CataII}.
At large $-k^2$ these can be computed perturbatively in QCD and one finds
\begin{align} \label{qcdres}
& \Pi^{V,V}(k^2) \rightarrow  \frac{N_c}{24 \pi^2} \log(-k^2)~,\cr
& \Pi^{T,V}(k^2) \rightarrow  - \frac{N_c}{4\pi^2} m_q \log{(-k^2)}~, \cr
& \Pi^{T,T \parallel}(k^2) \rightarrow  \frac{N_c}{24 \pi^2} \log(-k^2)~, \qquad \Pi^{T,T \perp}(k^2) \rightarrow - \frac{N_c}{24 \pi^2}  \log(-k^2)~,
\end{align}
where $m_q$ is the quark mass.  In the chiral limit, $m_q \to 0$,
the tensor and vector currents do not mix perturbatively. This is
consistent with the observation that $O^T,O^V$ transform in
different representations of the chiral symmetry group. Of course
the perturbative mixing is proportional to the explicit chiral
symmetry breaking parameter -- but one should expect nonperturbative
mixing, even in the chiral limit, due to the spontaneous chiral
symmetry breaking of the quark condensate.

Given \eqref{Oodd}-\eqref{OVrho}, it is clear that $\Pi^{V,V}$ and $\Pi^{T,T\parallel}$ should
feature resonances corresponding to $\omega/\rho$ exchange, while $\Pi^{T,T\perp}$ should have resonances
corresponding to $h_1/b_1$ exchange.  By inserting a complete set of states into the two-point functions one learns
\begin{align} \label{MEpoles}
& \Pi^{V,V}(k)  |_{\rm poles} =  -\sum_{n} \frac{  (f_{\rho}^{V,(n)} )^2 }{k^2 - (m_{\rho}^{(n)})^2}~, \qquad
\Pi^{T,T\parallel}(k) |_{\rm poles} = - \sum_{n} \frac{ (f_{\rho}^{T,(n)})^2 }{k^2 - (m_{\rho}^{(n)})^2 }~, \cr
& \Pi^{T,T\perp}(k)  |_{\rm poles}  = \sum_{n} \frac{ (f_{b}^{(n)})^2 }{k^2 - (m_{b}^{(n)})^2}~.
\end{align}
The decay ``constants'' here are the same as those appearing in \eqref{Oodd}-\eqref{OVrho}.  In general they run with scale.

\subsection{Low-lying hadron spectrum and decay constants}

As mentioned in the previous section, we can think of these vector and tensor operators as generating towers of increasingly
massive spin-one resonances from the vacuum.
The PDG summary table \cite{pdg} lists three spin-one mesons with the quantum numbers of the isotriplet $\rho$ meson:
$\rho(770)$, $\rho(1450)$, and $\rho(1700)$ with masses in ${\rm MeV}$ of $775.49 \pm 0.34 $, $1465 \pm 25 $, and $1720 \pm 20$
respectively; and three spin one mesons with the quantum numbers of the isosinglet $\omega$:
$\omega(782)$, $\omega(1420)$,  and $\omega(1650)$ with masses of $782.65 \pm 0.12 $, $1400-1450$, and $1670 \pm 30$.
Additional states such as the $\rho(1900)$, $\rho(2150)$ are mentioned in the complete particle listings
and their mass spacing was quoted in \cite{softwall} as part of the evidence for linearity in $n$ of the
mass squared of exited meson states $m_n^2$, but their existence must be regarded as uncertain. In comparison, the
experimental evidence for the three lightest $\rho$ states
is now quite compelling \cite{belle}.

 \begin{table}[htbp]
  \centering
  \begin{tabular}{@{} ccc @{}}
    \hline
    quantity & exp/lattice  result & source \\
    \hline\hline

                        $m_{\rho^0}$ & $775.49 \pm 0.34$  & \cite{pdg} \\
    $m_{\rho'}$ & $1465 \pm 25$ & \cite{pdg} \\
    $m_{\rho''}$ & $1720 \pm 20$ & \cite{pdg} \\
         $f_{\rho}$& $153 \pm 7$ & \cite{Donoghue} \\
          $f_{\rho'}$&  N/A&  \\
    $f_{\rho''}$ & N/A&  \\
      $f_{\rho}^{T}$  & ~$184 \pm 15$ & \cite{lattb1two} \\
        $f_{\rho'}^{T}$  & N/A&  \\
          $f_{\rho''}^{T}$  & N/A&  \\
          \hline
          $m_\omega$ & $872.65\pm0.12$ & \cite{pdg} \\
          $m_\omega'$ & $1400-1450$ & \cite{pdg} \\
       \hline
                          $m_{b_1}$ & $1229.5 \pm 3.2$ & \cite{pdg} \\
                              $f_{b_1}$ & $236 \pm 23$& \cite{latticeb1} \\
                          \hline
                          $m_{h_1}$ & $1170\pm 20$ & \cite{pdg} \\
                          \hline
                         $m_{\pi^0}$ & $134.9766\pm 0.0006$ & \cite{pdg} \\
                         $f_\pi$ & $92.4\pm 0.35$ & \cite{pdg} \\
                         $m_{a_1}$ & $1320\pm 40$ & \cite{pdg} \\
                         $f_{a_1}$ & $433\pm 13$ & \cite{pdg} \\
                         $m_{f_1}$ & $1281.8\pm 0.6$ & \cite{pdg} \\
  \end{tabular}
  \caption{Masses and decay constants of low-lying mesons in MeV.  The decay constants $f_{b_1}$, $f_\rho^T$ are
  evaluated at a scale of $2 ~ \textrm{GeV}$. Our focus is on  $1^{--}$ and $1^{+-}$ states; the lightest axial-vectors
  and pseudoscalars are included for completeness. }
  \label{table1}

\end{table}

We will denote the lowest three mass eigenstates by $\rho$, $\rho'$
and $\rho''$.  The corresponding vector and tensor decay constants,
as defined by \eqref{Oodd}-\eqref{OVrho}, will be denoted
$f_{\rho},f_{\rho'},f_{\rho''}$ and $f_{\rho}^T, f_{\rho'}^T,
f_{\rho''}^T$.  In table \ref{table1}
we summarize experimental results for masses and lattice results for
decay constants of the tensor and vector mesons.

\section{Holographic dual at the free level}

Having reviewed the experimental and lattice data on the QCD side,
we now extract this data from a holographic model. We first identify
the field dual to the tensor operator $O^T$ and its Lagrangian on
the hard wall background. We then compute the masses and decay
constants from the free equations of motion.

\subsection{Field-operator correspondence and quadratic action}

The gravitational background of the hard wall model consists of
$AdS_5$ truncated at a finite radius \cite{hardwall}. Working with
coordinates $x^M= (x^\mu,z)$ we use the metric
\beq \label{AdSmet}
ds^2 = \frac{\ell^2}{z^2} (\eta_{\mu \nu} dx^\mu dx^\nu -
dz^2)~, \qquad  \varepsilon \le z \le z_0~, \eeq
where $\ell$ is the $AdS$ radius and $z_0$ the infrared (IR) cutoff.
(The $AdS$ radius is often set to one, but we find it both useful and
less confusing to keep it explicit). We will perform calculations at
finite $\varepsilon$, taking the limit $\varepsilon \rightarrow 0$
at the end.

In what follows, $e^{MNPQR}$ will denote the 5d Levi-Civita tensor
with normalization $e^{0123z}=1/\sqrt{g}$. We will often rewrite
parts of the 5d action in terms of 4d fields; when we do so Greek
indices $\mu, \nu, \cdots$ will be raised and lowered with the flat
Minkowski metric $\eta_{\mu \nu}$. We also use $\epsilon^{\mu \nu
\lambda \rho}$ for the 4d Levi-Civita symbol with
$\epsilon^{0123}=1$.

The hard wall model of \cite{hardwall} introduces the fields dual to
the dimension three operators $O^{S,P}$ and to  $O^{V,A}$ in the
above background. These are, respectively, a tachyon field $X$  in
the bifundamental of the $U(N_f)_L \times U(N_f)_R$ gauge group, and
the (axial)-vector gauge fields $A$, $V$. The fields live in a
background geometry \eqref{AdSmet}.

We now extend the hard wall model to include a field dual to the
tensor operator $O^T$.  The obvious choice is a two-form potential
$b_{MN}$. To match the global flavor transformation properties of
$O^T$, $b_{MN}$ must transform as a bifundamental under the
$U(N_f)_L \times U(N_f)_R$ gauge group, and must therefore be a
complex field.

In order to match the six physical degrees of freedom generated by
$O^T$ (3 for each of the two massive vector-like states), $b_{MN}$
should also carry a total of six degrees of freedom on shell. The
complex field $b_{MN}$ has twenty independent real components, but as
will be seen explicitly later, the equations of motion imply a
Proca-like condition which removes eight of these, leaving twelve
independent on-shell degrees of freedom -- twice as many as we want.
In order to eliminate these additional states, one might consider
imposing a tensor gauge invariance, but this turns out to be
impossible. Since $b$ transforms nontrivially under the gauge group,
the usual three-form field strength $H=db$ is not gauge covariant.
On the other hand, the gauge covariant antisymmetrized derivative
\beq
{\mathcal H}_{PMN}= 3 D_{[P} b_{MN]}
\eeq
is not invariant under tensor gauge transformations $\delta b_{MN} =
2\partial_{[M} \lambda_{N]}$ and as a result one cannot use tensor gauge
invariance to remove the unphysical degrees of freedom.

This problem is familiar from the AdS/CFT literature where one also
finds two-form tensor fields transforming under the $SO(6)$ gauge
group in the dimensional reduction of IIB string theory on $S^5$
\cite{tobeI,tobeII}. These fields are described by a first-order
Lagrangian discussed in \cite{Townsend} and have been analyzed in
the context of the AdS/CFT correspondence in
\cite{AFII,Minces}\footnote{At the level of free field theory one
can establish an equivalence between the first order formalism and a
second order formalism with a conventional kinetic energy term and a
Chern-Simons mass term as in \cite{Pernici} but is is not clear that
this equivalence can be established when interactions are
included.}.  Though there is no tensor gauge invariance, none is
needed since a first-order action has fewer independent degrees of
freedom. In a first-order formalism half of these are momenta and
half are coordinates, so one is left with six independent real
degrees of freedom. This correctly matches the six physical degrees
of freedom (two massive spin one vectors) created by the dual
operator $O^T$.

According to the AdS/CFT map, a $d/2$-form field in $AdS_{d+1}$
obeying a first-order equation with mass $\mu$ (in units of the $AdS$
radius) is dual to an operator with dimension
\beq
\Delta= \frac{1}{2} (d+2 |\mu|)~.
\eeq
In contrast to the scaling dimension of the conserved current $J_V$,
we should expect corrections to the naive value of $\Delta=3$ for
$O^T$. We thus leave $\mu$ arbitrary.

We now have the following extended version of the hard wall action:
\beq \label{snought}
S = S_{\rm hw}+ S_{\rm CS}+S_{\rm sd}  +S_{\rm int}~,
\eeq
where
\beq \label{shw}
S_{\rm hw}= \int d^5x \sqrt{g} \  \tr \ \left\{ |DX|^2 + \frac{3}{\ell^2} |X|^2 - \frac{1}{4 \ell g_5^2} (F_L^2 +F_R^2) \right\}
\eeq
is the action used in \cite{hardwall}, and
\beq \label{scs}
S_{\rm CS} =  \frac{N_c}{24 \pi^2 } \int_{\mathcal{M}} \biggl( \omega_5(A_L) - \omega_5(A_R) \biggr)~,
\eeq
with $\tr F^3 = d \omega_5$, is the Chern-Simons term needed to match the flavor anomalies of
QCD (see e.g. \cite{Domokos}), and
\begin{align} \label{firstorderaction}
S_{\rm sd} = -  \frac{i }{2\ell g_{b}^2}\int_{\cM} \tr \ \left[
\bar{b}\left( D-i \frac{\mu}{\ell} \star\right)b -b\left(
D+i\frac{\mu}{\ell}\star\right)\bar{b}\right]  - \frac{{\rm
sgn}(\mu) }{4\ell g_b^2}\int_{ \dt \cM}\tr \ \bb_{\mu\nu}
b^{\mu\nu}~
\end{align}
is the first order action for the antisymmetric tensor field written in terms of differential forms\footnote{We have
$b = \frac{1}{2} b_{MN} dx^M dx^N$ and $Db \equiv \frac{1}{2} D_{[p} b_{MN]} dx^P dx^M dx^N$, and
the Hodge dual is $(\star b)_{MNP} = \frac{1}{2} e_{MNP}^{\phantom{MNP}QR} b_{QR}$.  Since $b$ is
a bifundamental, like $X$, the gauge covariant derivative acts as $D_P b_{MN} = \dt_P b_{MN} - i A_{L,P} b_{MN} + i b_{MN} A_{R,P}$.}.
We use a bar to denote Hermitian conjugation on group indices--e.g. $\overline{X} \equiv X^\dag$.  Since we work in a basis of Hermitian generators, this amounts to complex conjugation of the individual flavor components.
The interaction term $S_{\rm int}$ will be discussed
later. Note that we have explicitly inserted factors of the $AdS_5$ radius, $\ell$, in such a way that the
couplings $g_5$ and $g_b$ are dimensionless.  The mass dimensions of the fields are $[X] = [b] = 3/2$, $[A_{L,R}] = 1$.

First order actions on manifolds with boundary require the addition
of boundary terms as dictated by the requirement of a consistent
variational principle \cite{Henneaux} or from consistency when
passing between the Lagrangian and Hamiltonian formulations of the
theory \cite{AFI}.  These terms play a crucial role in giving a
precise definition to the AdS/CFT correspondence since the bulk
action vanishes on the equations of motion, leaving the boundary
term to generate correlation functions when we vary with respect to
the sources.

The boundary term in \eqref{firstorderaction} was first found by
\cite{AFII}, but we can also obtain it via a variational argument as
in \cite{Henneaux}.  Since this term will play an important role, we
review the argument here.  The boundary term is required for the
variational problem to be well-defined: without the boundary term
the variation of the action evaluated on shell will not vanish,
implying that solutions to the equations of motion do not represent
stationary points of the action, and thus invalidating the
stationary phase approximation to the partition function.

We begin by analyzing the equations of motion to determine the appropriate degrees of freedom.
Taking the variation of $S_{\rm sd}$ with respect to $\bar{b}$, the (free) equations of motion are
\begin{align}\label{beom}
\left(d-i\frac{\mu}{\ell}\star\right)b =0~.
\end{align}
We review the solution to these equations in some detail in Appendix
\ref{ap:gensol}.  Here we note simply that $b_{\mu z}$ is determined
in terms of $b_{\mu\nu}$, and one may derive separate equations for
the self-dual and anti-self-dual parts of $b_{\mu\nu}$.  Writing
\begin{align}
b_{\mu\nu}(x,z)=b^+_{\mu\nu}(x,z)+b^-_{\mu\nu}(x,z)\quad\textrm{with}\quad b^{\pm}_{\mu\nu}=\pm\frac{i}{2}\epsilon_{\mu\nu\rho\sigma}b^{\pm\rho\sigma}~,
\end{align}
the (anti) self-dual pieces obey the equations
 \begin{align}\label{bpmeoms}
 \left[ \partial^2_z-\frac{1}{z}\partial_z+k^2-\frac{\mu(\mu\pm2)}{z^2} \right]b^{\pm}_{\mu\nu}(k,z)=0~,
 \end{align}
where $b(k,z) = \int d^4 x e^{i k \cdot x} b(x,z)$, as usual.
Near the UV boundary $z\sim\varepsilon \to 0$, solutions to \eqref{bpmeoms} behave as
\begin{align}
& b^+_{\mu\nu}\sim \tilde{S}_{\mu\nu}\varepsilon^{-\mu} - \tilde{s}_{\mu\nu} \varepsilon^{2+\mu}~, \\
& b^-_{\mu\nu}\sim \tilde{A}_{\mu\nu}\varepsilon^{2-\mu}-\tilde{a}_{\mu\nu}\varepsilon^{\mu}~,
\end{align}
where $\tilde{S},\tilde{s}$ and $\tilde{A},\tilde{a}$ are self-dual and anti-self-dual polarizations, respectively.
Since the equation of motion \eqref{beom} is first order, however, these coefficients are not independent.  One may derive the relation
\begin{align}\label{polrel}
& (\tilde{A},\tilde{a})_{\mu\nu} = \frac{1}{k^2} \left( (\mathcal{P}^\perp)_{\mu\nu}^{\alpha\beta} - (\mathcal{P}^\parallel)_{\mu\nu}^{\alpha\beta} \right) (\tilde{S}, \tilde{s} )_{\alpha\beta}~.
\end{align}
Equivalently, we have
\begin{align}\label{polrelalt}
& (\tilde{S},\tilde{s})_{\mu\nu} =\frac{1}{k^2} \left( (\mathcal{P}^\perp)_{\mu\nu}^{\alpha\beta} - (\mathcal{P}^\parallel)_{\mu\nu}^{\alpha\beta} \right) (\tilde{A}, \tilde{a} )_{\alpha\beta}~,
\end{align}
which is consistent with \eqref{polrel} since $(\mathcal{P}^\perp - \mathcal{P}^\parallel)^2 = k^4 {\bf 1}$.  Observe that when $\mu >0$, $\tilde{S}$ encodes the leading behavior of $b_{\mu\nu}$ near the UV boundary, while when $\mu < 0$, $\tilde{a}$ encodes the leading behavior.  It will be important to distinguish these two cases below.

Now consider the variation of the bulk part of
$S_{\rm sd}$, focusing on terms with $z$-derivatives:
\begin{align}
\delta S_{\rm sd}^{\rm bulk} =&~ - \frac{i}{8\ell g_{b}^2} \delta \int_{\cM} d^5 x \epsilon^{\mu\nu\rho\sigma} {\rm tr} \left\{ \bar{b}_{\mu\nu} \dt_z b_{\rho\sigma} - b_{\mu\nu} \dt_z \bar{b}_{\rho\sigma} \right\} + \cdots \nonumber \\
=&~ - \frac{i}{8\ell g_{b}^2} \delta \int_{\cM} d^5 x \epsilon^{\mu\nu\rho\sigma} {\rm tr} \left\{ \overline{b_{\mu\nu}^+} \dt_z b_{\rho\sigma}^- + \overline{b_{\mu\nu}^-} \dt_z b_{\rho\sigma}^+ - b_{\mu\nu}^+ \dt_z \overline{b_{\rho\sigma}^-} - b_{\mu\nu}^- \dt_z \overline{b_{\rho\sigma}^+} \right\} + \cdots  \nonumber \\
=&~ - \frac{i}{4\ell g_{b}^2} \int_{\cM} d^5 x \epsilon^{\mu\nu\rho\sigma} {\rm tr} \left\{ \dt_z b_{\mu\nu}^- \delta \overline{ b_{\rho\sigma}^+} + \dt_z b_{\mu\nu}^+ \delta \overline{ b_{\rho\sigma}^-} - \dt_z \overline{b_{\mu\nu}^-} \delta b_{\rho\sigma}^+ - \dt_z \overline{b_{\mu\nu}^+} \delta b_{\rho\sigma}^- \right\} + \cdots \nonumber \\
&~ \quad - \frac{i}{8\ell g_{b}^2} \int d^4 x \epsilon^{\mu\nu\rho\sigma} {\rm tr} \left\{ \overline{b_{\mu\nu}^+} \delta b_{\rho\sigma}^- + \overline{b_{\mu\nu}^-} \delta b_{\rho\sigma}^+ - b_{\mu\nu}^+ \delta \overline{b_{\rho\sigma}^-} - b_{\mu\nu}^- \delta \overline{b_{\rho\sigma}^+} \right\}_{\varepsilon}^{z_0}~. \label{Ssdbulkvary}
\end{align}
Evaluating the variation on shell, the bulk terms vanish by the
equations of motion. The relation \eqref{polrel} implies that we
cannot simultaneously fix
 $b_{\mu\nu}^+$ and $b_{\mu\nu}^-$ at either the UV boundary or the
IR boundary--to do so would overconstrain the system.  At each
boundary we may fix only one or the other.  The natural choice at
the UV boundary, from the AdS/CFT point of view, is to fix $b^+$ $(b^-)$ for $\mu >0$ $(\mu < 0)$.
(In the language of \cite{AFII}, for $\mu >0$, $b^+$ plays the role of coordinate
and $b^-$  that of canonical momentum, while for $\mu <0$ their roles are reversed).  Either way, half of the boundary
terms in the last line of \eqref{Ssdbulkvary} will remain. If we constrain $b^+$, say, then the
$\delta b^-$ terms will be nonzero.

There is a way out of this quandary.  First let us suppose $\mu >0$.  Notice that the on-shell value of
\eqref{Ssdbulkvary} may be written as
\begin{align}
\delta S_{\rm sd}^{\rm bulk} =&~ - \frac{i}{8\ell g_{b}^2} \delta \int d^4 x \epsilon^{\mu\nu\rho\sigma} {\rm tr} \left\{ \overline{b_{\mu\nu}^+} b_{\rho\sigma}^- - b_{\mu\nu}^+ \overline{b_{\rho\sigma}^-} \right\}_{\varepsilon}^{z_0} + \nonumber \\
&~  \qquad \qquad \qquad + \frac{i}{4 \ell g_{b}^2} \int d^4 x \epsilon^{\mu\nu\rho\sigma} {\rm tr} \left\{ b_{\mu\nu}^- \delta \overline{b_{\rho\sigma}^+} - \overline{b_{\mu\nu}^-} \delta b_{\rho\sigma}^+ \right\}_{\varepsilon}^{z_0}~,
\end{align}
or, moving the first term to the left and using the (anti) self-duality of $b^{\pm}$,
\begin{align}
& \delta \left( S_{\rm sd}^{\rm bulk} - \frac{1}{4\ell g_{b}^2} \int_{\dt \cM} {\rm tr} \left\{ \bar{b}^{\mu\nu} b_{\mu\nu} \right\} \right) =  \frac{i}{4 \ell g_{b}^2} \int d^4 x \epsilon^{\mu\nu\rho\sigma} {\rm tr} \left\{ b_{\mu\nu}^- \delta \overline{b_{\rho\sigma}^+} - \overline{b_{\mu\nu}^-} \delta b_{\rho\sigma}^+ \right\}_{\varepsilon}^{z_0}~.
\end{align}
If $\mu < 0$ we should instead isolate $\delta b^-$.  We find that the on-shell value of \eqref{Ssdbulkvary} may also be rewritten as
\begin{align}
& \delta \left( S_{\rm sd}^{\rm bulk} + \frac{1}{4\ell g_{b}^2} \int_{\dt \cM} {\rm tr} \left\{ \bar{b}^{\mu\nu} b_{\mu\nu} \right\} \right) =  \frac{i}{4 \ell g_{b}^2} \int d^4 x \epsilon^{\mu\nu\rho\sigma} {\rm tr} \left\{ b_{\mu\nu}^+ \delta \overline{b_{\rho\sigma}^-} - \overline{b_{\mu\nu}^+} \delta b_{\rho\sigma}^- \right\}_{\varepsilon}^{z_0}~.
\end{align}
Thus, by adding the boundary term, as has been done in
\eqref{firstorderaction}, we allow for a consistent variational
principle. The on-shell variation of the full (bulk $+$ boundary)
action will vanish if the boundary value of $b^+$ $(b^-)$ is held fixed in the
case $\mu >0$ $(\mu < 0)$.

We in fact have two choices for the boundary condition. Consider the $\mu >0$ case for concreteness;
analogous comments apply in the $\mu < 0$ case.  When $\mu > 0$, the on-shell variation of \eqref{firstorderaction} will vanish if we either hold $b^+$ fixed \textit{or} set
$b^- = 0$ on the boundary.  We will refer to these as the
Dirichlet-type and Neumann-type conditions, respectively, drawing
on the coordinate and momentum characterization of $b^{\pm}$
given in \cite{AFII}.   On the UV boundary
AdS/CFT dictates that we take the Dirichlet-type condition, holding $b^+$
fixed, since $b^+$ represents the source for the dual operator.  On the
IR boundary, the correct choice is not immediately obvious.

The question is compounded by the potential presence of higher-order
terms localized on the IR boundary and not included in \eqref{snought}.
In general we should expect such terms to appear as a result of
``integrating out'' nontrivial IR dynamics up to the scale of the IR
cutoff $1/z_0$. These terms could potentially modify the
Neumann-type condition to a more general mixed condition,
which may even be nonlinear, with parameters depending on the
details of the localized terms\footnote{We take the point of view
that effects of higher order boundary terms modify the boundary
conditions, such that a well defined variational principle is
maintained.  An alternative approach is to hold the simple
Neumann-type boundary condition fixed, so the presence of
higher-order terms leads to a contribution to the on-shell
\emph{value} of the action from the IR boundary.  On a practical
level one may take either point of view; these are just two
different ways of encoding the effects of such higher-order terms.}.

These issues were dealt with in two different ways in the original
hard wall model \cite{hardwall}, depending on the dual field in
question.  In the case of the scalar field, $X$, all of these
unknowns were packaged into a single parameter, the quark
condensate, which was taken as an input that could be fit to data.
In the case of the vector gauge field, $V$, the simple Neumann
condition, $F_{z\mu}(z_0) \sim \d_z V_\mu |_{z_0} = 0$, was chosen,
which can be motivated as follows. First, one can definitively
choose in favor of the Neumann condition over the Dirichlet one,
$\delta V |_{z_0} = 0$, by requiring that the boundary condition be
gauge invariant. This does not rule out the possibility of higher
order terms in the field strength localized at the IR boundary, but
the leading term of this sort was considered, and its effect was
found to be small in practice.  In general, one expects terms
involving higher powers of the gauge field--on the boundary or in
the bulk--to be suppressed by powers of $1/N_c$, while for $X$ this
is not true.  We will review these arguments in section 4.

Returning to the case at hand, one may also use $N_c$ counting to argue that higher order
terms in $b$ are suppressed.  We do not have an analog of the gauge principle, however, to help us
decide between the Neumann or Dirichlet-like condition.  In the next subsection
we use the holographic model \eqref{snought} to compute the tensor-tensor two-point
function, $\langle O^T O^T \rangle$, considering both IR boundary conditions.  We will show that
there is a strong physical argument for choosing the Neumann-like condition  over
the Dirichlet one: the Dirichlet-like condition leads to a massless divergence in the tensor-tensor two-point
 function, and there are no massless particles with the quantum numbers of the $h_1/b_1$-mesons in QCD!
 The Neumann-like condition, on the other hand, gives a sensible finite result for $\langle O^T O^T \rangle$
 in the $k \to 0$ limit.

\subsection{On-shell action and tensor-tensor two-point function}

Let us now use the quadratic action \eqref{snought}, and in particular \eqref{firstorderaction}, to study
the correlation functions and the normalizable spectrum of the complex two-form $b_{MN}$.
The two-form $b_{MN}$ corresponds to a total of six real degrees of freedom.  We can package these in a variety of ways: in terms of real and imaginary, longitudinal and transverse,
or self-dual and anti-dual parts of $b_{\mu\nu}$\footnote{The components $b_{\mu z}$ can be eliminated through their equations of motion.}. It will therefore be convenient to work as much as possible in terms of  projections onto these parts, $P^{\pm}$ and
$\mathcal{P}^{\parallel,\perp}$,  defined in Appendix \ref {ap:proj}.
The intermediate steps of the analysis differ between the cases $\mu >0$ and $\mu <0$, though the final result for the tensor-tensor two-point function can be given in a simple form, valid for both cases.  In order to streamline the discussion in this subsection, we will present the $\mu > 0$ case only, and at the end quote the result for general $\mu$.  Details of the analysis are relegated to appendices \ref{ap:gensol} and \ref{ap:MEcomp}.

We first determine the bulk-to-boundary propagator of $b$, which we need in order to compute
two-point functions of the dual operators.

For the free two-form in $AdS_5$ it is most convenient to work in terms of the self-dual piece which satisfies
\begin{align}
b^+_{\mu\nu}=\frac{i}{2}\epsilon_{\mu\nu}^{\phantom{\mu\nu}\rho\sigma}b^+_{\rho\sigma}~.
\end{align}
As discussed by \cite{AFII}, it is the self-dual piece which sources the dual operator: for the value $\mu=1$, which
should give the scaling dimension in a \emph{conformal} theory (i.e. on a background that is simply $AdS_5$),
it indeed has the appropriate near-boundary scaling behavior to correspond to a tensor operator. In the confining theory, of course,
the value of $\mu$ should receive quantum corrections.

The general solution to the free equations of motion for $b$ is
reviewed in Appendix \ref{ap:gensol}. In momentum space one has
$b_{\mu\nu} = b_{\mu\nu}^+ + b_{\mu\nu}^-$, with
\begin{align}\label{bgensols}
&  b_{\mu\nu}^+(k,z) = \tilde{S}_{\mu\nu}(k) z J_{-\mu-1}(k z) + \tilde{s}_{\mu\nu}(k) z J_{\mu +1}(k z)~, \nonumber \\
& b_{\mu\nu}^-(k,z) = \tilde{A}_{\mu\nu}(k) z J_{-\mu+1}(k z) + \tilde{a}_{\mu\nu}(k) z J_{\mu -1}(k z)~,
\end{align}
where the anti-self-dual polarizations are related to the self-dual
ones by
\begin{align}
& \tilde{A}_{\mu\nu} = \frac{1}{k^2} (\mathcal{P}^\perp - \mathcal{P}^\parallel)_{\mu\nu}^{\alpha\beta} \tilde{S}_{\alpha\beta} = \tilde{S}_{\mu\nu} - \frac{2}{k^2} \left( k_\mu k^\rho \tilde{S}_{\rho\nu} - k_\nu k^\rho \tilde{S}_{\rho\mu} \right)~,
\end{align}
and similarly for $\tilde{a}$ in terms of $\tilde{s}$.  Meanwhile,
$b_{\mu z}$, which plays the role of Lagrange multiplier, is given
by $b_{\mu z} = -\frac{z}{2\mu}
\epsilon_{\mu}^{\phantom{\mu}\nu\rho\sigma} k_\nu b_{\rho\sigma}$.
The solution \eqref{bgensols} is appropriate in the generic case of
non-integer $\mu$; if $\mu$ is an integer then the Bessel functions
$J_{-\mu \mp1}$ should be replaced by $Y_{\mu \pm 1}$.

We are interested specifically in  the bulk-to-boundary propagator:
that is, the solution to the equations of motion with the boundary condition that
the self-dual field approaches a self-dual source
on the UV boundary.

We first impose an IR boundary condition at $z = z_0$ to fix
$\tilde{s}$ in terms of $\tilde{S}$. As we discussed above, there
are two possibilities for a consistent variational principle: we may
choose the Dirichlet-like condition where we hold $\delta b^+$
fixed, or the Neumann-like condition where we set $b_{\mu\nu}^-(z_0)
= 0$.  Holding $b^+$ fixed in this context means setting
$b_{\mu\nu}^+(z_0) = 0$, since there is no natural constant
antisymmetric two-tensor living on the IR boundary.  After imposing
one of these, we find that the solution takes the form
\begin{align}\label{IRimposed}
& b_{\mu\nu}^+(k,z) = \tilde{S}_{\mu\nu}(k) \left[ z J_{-\mu-1}(k z) - c_{b}^{>}(k,z_0) z J_{\mu +1}(k z) \right] \equiv \tilde{S}_{\mu\nu}(k) B_{>}^+(k,z)~, \nonumber \\
& b_{\mu\nu}^-(k,z) = \tilde{A}_{\mu\nu}(k) \left[ z J_{-\mu+1}(k z) - c_{b}^{>}(k,z_0) z J_{\mu -1}(k z) \right] \equiv \tilde{A}_{\mu\nu}(k) B_{>}^-(k,z)~,
\end{align}
where
\begin{align}\label{cbdef}
c_{b}^{>}(k,z_0) = \left\{ \begin{array}{c} \frac{J_{-\mu-1}(k z_0)}{J_{\mu +1}(k z_0)} ~, \qquad b^+(z_0) = 0~, \cr\cr
\frac{J_{-\mu +1}(k z_0)}{J_{\mu -1}(k z_0)} ~, \qquad b^-(z_0) = 0~. \end{array} \right.
\end{align}
(The ``$>$'' labels are a reminder that these expressions are appropriate for the case $\mu > 0$).

The UV boundary condition fixes $\tilde{S}_{\mu\nu}$ in terms of the
source for the dual operator. As we approach the UV boundary, $z =
\varepsilon \to 0$, the leading scaling behavior of $b$ is $b^+
\propto \varepsilon^{-\mu}$.  We define the self-dual source,
$S_{\mu\nu}(k)$, through
\begin{align}\label{UVimposed}
b_{\mu\nu}^+(k,\varepsilon) = \frac{\ell^{\mu -1/2}}{\varepsilon^\mu} S_{\mu\nu}(k)~.
\end{align}
The factors of the $AdS$ radius $\ell$ have been inserted on dimensional grounds.  $b$ is a
dimension $3/2$ field, while $S$ should have naive dimension 1 since it sources a naive dimension 3 operator.
Using this boundary condition to eliminate $\tilde{S}$ in favor of the source $S$, we arrive at our final expression
for the bulk to boundary propagator:
\begin{align}\label{btob}
& b_{\mu\nu}^+(k,z) =   \ell^{\mu -1/2} S_{\mu\nu}(k) \frac{B_{>}^+(k,z)}{\varepsilon^\mu B_{>}^+(k,\varepsilon)}~, \nonumber \\
& b_{\mu\nu}^-(k,z) = \ell^{\mu -1/2} \left[ S_{\mu\nu} - \frac{2}{k^2} \left( k_\mu k^\rho S_{\rho\nu} - k_\nu k^\rho S_{\rho\mu} \right) \right] \frac{B_{>}^-(k,z)}{\varepsilon^{\mu} B_{>}^+(k,\varepsilon)}~.
\end{align}

Pulling out the explicit factor of $\varepsilon^{-\mu}$, as we have
done in \eqref{UVimposed}, ensures that our final expressions for
two-point functions will be $\varepsilon$-independent and
corresponds to working with ``rescaled'' operators.  Another
procedure commonly appearing in the literature is to simply set
$\ell^{1/2} b^+(k,\varepsilon) = S(k)$, in which case our result for
the two-point function would include an overall factor of
$\varepsilon^{2\mu}$, indicating its scaling behavior as one
approaches the UV boundary.  In the $\varepsilon \to 0$ limit one
should trade this factor for a renormalization scale $M_r$. In
\eqref{UVimposed} we have chosen to identify the renormalization
scale with the $AdS$ radius, $M_r \sim \ell^{-1}$. This is natural
since the dimensionful couplings $\ell^{1/2} g_5$, $\ell^{1/2}
g_{b}$ appearing in \eqref{snought} were already (implicitly)
defined at this scale.

To summarize, three different length scales appear in the model: the
locations of the UV and IR boundaries at $\varepsilon$ and $z_0$,
respectively, and $\ell$ the $AdS$ radius. In terms of the dual
field theory, the mass scale $\varepsilon^{-1}$ is the UV cutoff we
would include in the computation of bare $n$-point functions. Adding
the appropriate counterterms to the action is essentially equivalent
to replacing $\varepsilon^{-1}$ with $\ell^{-1}$, to give a finite
(renormalized) result. $\ell^{-1}$ is the renormalization scale,
while $z_0^{-1}$ is $\Lambda_{\rm QCD}$.

These statements may seem trivial when we are working with a
simplified model consisting of $AdS_5$ with cutoff.  While this model captures basic features such as confinement, couplings and dimensions of operators do not run with scale. In a more realistic model this running would be encoded in a nontrivial $z$-dependent geometry. Physically, the picture is that the asymptotic part of our $AdS$
slice is not describing asymptotically free QCD, but  rather is
providing an approximate description in a window of scales where the
QCD coupling is finite, but running slowly with scale. 
We have already stressed that the tensor operator $O^T$ is not a
conserved current, and at strong coupling one should expect $\cO(1)$
corrections to its charge and dimension -- or to $g_b,\mu$ in the
dual language. The renormalization scale $\ell^{-1}$ represents a typical scale in this window. The ratio of this scale to the QCD scale,
parameterized by the ratio $z_0/\ell$, is a dimensionless parameter in our
model, and we will see that it naturally appears in physical
quantities like decay constants.

To compute correlators according to the usual AdS/CFT description,
we evaluate the action on the solution \eqref{btob}, functionally
differentiate with respect to the source, and then take the
$\varepsilon \to 0$ limit. $S_{\mu\nu}$ couples to the self-dual
operator, $O_{\mu\nu}^{T,+}$ in \eqref{Oasd}.  It is related to the
source, $\cT_{\mu\nu}$, for the tensor operator $O^T$ via
\begin{align}\label{sourcerlns}
& S_{\mu\nu} = \cT_{\mu\nu} + \frac{i}{2} \epsilon_{\mu\nu}^{\phantom{\mu\nu}\rho\sigma} \cT_{\rho\sigma} \quad \Rightarrow \quad \cT_{\mu\nu}(k) = \frac{1}{2} \left( S_{\mu\nu}(k) + \bar{S}_{\mu\nu}(-k) \right)~.
\end{align}
The sole contribution to the on-shell action comes from the UV boundary term, which we find to be
\begin{align}\label{Ssdonshell}
& S_{\rm sd} = \frac{\ell^{2\mu-2}}{g_{b}^2} \int \frac{d^4 k}{(2\pi)^4} \frac{B_{>}^-(k,\varepsilon)}{k^2 \varepsilon^{2\mu} B_{>}^+(k,\varepsilon)} {\rm tr} \left\{ \cT^{\alpha\beta}(-k) \left[ \mathcal{P}^{\perp}_{\alpha\beta,\delta\gamma} - \mathcal{P}^{\parallel}_{\alpha\beta,\delta\gamma} \right] \cT^{\delta\gamma}(k) \right\}~.
\end{align}

This action leads to the following matrix elements.  We have that
$\Pi^{T,T\parallel} = - \Pi^{T,T\perp}$, as is evident from
\eqref{Ssdonshell}, and
\begin{align}\label{PiTTcb}
\Pi^{T,T \perp}(k) =\left\{ \begin{array}{l l} - \frac{\Gamma(-\mu)}{2^{2\mu-2} g_{b}^2 \Gamma(\mu)} c_{b}^{>}(k,z_0) (k \ell)^{2\mu-2} ~, & ~\textrm{$\mu > 0$ non-integer}~, \\ & \\
\frac{1}{2^{2\mu-2} g_{b}^2 \mu! (\mu-1)!} \left[ \pi c_{b}^{>}(k,z_0) - \log{( k^2 \ell^2)} \right] (k \ell)^{2\mu-2}~, & ~\textrm{$\mu > 0$ integer}~, \end{array} \right.
\end{align}
where $c_{b}^{>}(k,z_0)$ is given by \eqref{cbdef} in the non-integer $\mu$ case and
by $\eqref{cbdef}$ with $J_{-\mu \mp 1} \to Y_{\mu \pm 1}$
in the integer $\mu$ case.  Some of the intermediate steps involved in
obtaining \eqref{Ssdonshell}, \eqref{PiTTcb}, are presented in
Appendix \ref{ap:MEcomp}.

Let us consider the extreme IR limit of this result, $k \to 0$.  The tensor-tensor
correlator should be finite in this limit,
since there would be no particle interpretation in QCD for such a massless divergence.
The leading $k^2$ behavior of $c_{b}^{>}(k,z_0)$ is
(for both integer and non-integer $\mu$)
\begin{align}\label{cbsmallk}
\lim_{k \rightarrow 0} c_{b}^{>}(k,z_0) = \left\{ \begin{array}{l l} \cO(k^{-2\mu -2})~, & \quad b^+(z_0) = 0~, \\
\cO(k^{-2\mu + 2})~, & \quad b^-(z_0) = 0~. \end{array} \right.
\end{align}
We conclude that the Dirichlet-like boundary condition, $b^+(z_0) = 0$, is
physically unacceptable since it leads to a divergence in $\Pi^{T,T}$ at large distances.
The Neumann-like condition, $b^-(z_0) = 0$, on the other hand gives a physically reasonable result
for the $k \to 0$ limit of $\Pi^{T,T}$.  Per our discussion in the previous subsection, we take
the Neumann-like boundary condition to be our IR condition for the $b$-sector.

The final result for the tensor-tensor two-point function, in the case $\mu > 0$, computed using the free dual \eqref{firstorderaction}, is
\begin{align}\label{PiTTfinal}
\Pi^{T,T \perp}(k) =\left\{ \begin{array}{l l} - \frac{\Gamma(-\mu)}{2^{2\mu-2} g_{b}^2\Gamma(\mu)} (k \ell)^{2\mu-2} \frac{J_{-\mu+1}(k z_0)}{J_{\mu-1}(k,z_0)} ~, & ~\textrm{$\mu > 0$ non-integer}~, \\ & \\
\frac{1}{2^{2\mu-2} g_{b}^2 \mu! (\mu-1)!} (k \ell)^{2\mu-2} \left[  \pi \frac{Y_{\mu-1}(k z_0)}{J_{\mu-1}(k z_0)} - \log{( k^2 \ell^2)} \right] ~, & ~\textrm{$\mu > 0$ integer}~. \end{array} \right.
\end{align}
In appendices \ref{ap:gensol} and \ref{ap:MEcomp} we also analyze the $\mu < 0$ case.  The final result is identical to \eqref{PiTTfinal} with $\mu \to |\mu|$, up to a sign whose origin may be traced to the sign on the boundary term in \eqref{firstorderaction}.  Thus the general result, valid for any $\mu \neq 0$, is
\begin{align}\label{PiTTfinalgenmu}
\Pi^{T,T \perp}(k) =\left\{ \begin{array}{l l} - \frac{{\rm sgn}(\mu) \Gamma(-|\mu|)}{2^{2|\mu|-2} g_{b}^2 \Gamma(|\mu|)} (k \ell)^{2|\mu|-2} \frac{J_{-|\mu|+1}(k z_0)}{J_{|\mu|-1}(kz_0)} ~, & ~\textrm{$\mu$ non-integer}~, \\ & \\
\frac{{\rm sgn}(\mu)}{2^{2|\mu|-2} g_{b}^2 |\mu|! (|\mu|-1)!} (k \ell)^{2|\mu|-2} \left[  \pi \frac{Y_{|\mu|-1}(k z_0)}{J_{|\mu|-1}(k z_0)} - \log{( k^2 \ell^2)} \right] ~, & ~\textrm{$\mu$ integer}~. \end{array} \right.
\end{align}
Next we turn to a discussion of the physics contained in \eqref{PiTTfinalgenmu}.

\subsection{Interpretation of results}

The original hard-wall model depends on the parameters $z_0$, $m_q$, and $\sigma$, as well as $g_5$ which is expressed in
units of the $AdS$ radius $\ell$. Our posited extension of the hard-wall model contains new parameters $\mu$, $g_b$
as well as explicit dependence on $\ell$.
All physical quantities we
compute will be given in terms of these parameters, so in order to
make additional predictions, we need to fix them using real QCD
results. Of course it is best, when possible, to fix these
parameters exactly using explicit computations in QCD. For example,
quantities which are not renormalized, such as the anomalous dimension of a conserved current, give the same result at  weak or strong coupling. The
authors of \cite{hardwall} successfully employed this strategy to
find $g_5$ in \eqref{snought}, by comparing the gravity dual and
perturbative QCD results for the large $Q^2$ $J_V$-$J_V$ correlator.
Naively, one might want to fix $\mu$ and $g_b$ by comparing the UV
(large momentum) limit of the tensor two-point functions to
large-momentum QCD. While this method is justified for finding $g_5$
using the two-point function of the \textit{conserved} current
$J_V$, the operator $O^T$ is not conserved, so we have no reason to
expect that $\mu$ and $g_b$ will not receive significant corrections
at strong coupling. (Incidentally, the issue applies to the scaling
dimension of $\bar{q}q$, dual to the 5d mass of the tachyon field $X$ which was held
at its naive value in \cite{hardwall}).

The two-point function \eqref{PiTTfinalgenmu} encapsulates much of the new information we obtain
from adding the non-interacting two-form field: the locations of its $k$-space poles mark the masses
of mesonic resonances generated by $O^T$ while the residues at the poles give the corresponding decay constants. We can
expand the two-point function around its poles, and by comparison
to \eqref{MEpoles}, read off these quantities\footnote{This is equivalent to identifying the masses as the eigenvalues
of normalizable modes and the decay constants as derivatives of
the eigenfunctions. Especially when one studies interactions it is often useful to work with the normalizable eigenfunctions,
and to write down an effective 4d action for the corresponding
resonances by integrating out the holographic $z$ direction.
We consider this point of view in Appendix \ref{ap:normmodes}, but here it is equally convenient to just work with the two-point
function.}.

By inspection of \eqref{PiTTfinalgenmu} we see that the two-point function (for both integer and non-integer $\mu$) has an infinite set of
simple poles for values of $k=m$ such that $J_{|\mu|-1}(mz_0)=0$:  these define the masses of the states created
by $O^T$ to be $m_n=x_{|\mu|-1,n}z_0^{-1}$ where $x_{|\mu|-1,n}$ denotes the $n$-th zero of $J_{|\mu|-1}$. Note that each pole corresponds to
two degenerate states: a $1^{+-}$ ($h_1/b_1$-like) state
and a $1^{--}$ ($\omega/\rho$-like) state.

Taylor-expanding near the first pole, we find
\begin{align}
\Pi^{T,T\perp}=\left\{
\begin{array}{l}
-\frac{{\rm sgn}(\mu)}{g_b^2} \left( \frac{x_{|\mu|-1,1}\ell}{2z_0} \right)^{2|\mu|-2}\frac{2 \Gamma(-|\mu|) J_{1-|\mu|}(x_{|\mu|-1,1})}{x_{|\mu|-1,1} \Gamma(|\mu|) J'_{|\mu|-1}(x_{|\mu|-1,1})} \ \frac{m_1^2}{k^2-m_1^2}+\cdots~,  \\
\qquad \qquad \qquad \qquad \qquad \qquad \qquad \qquad \qquad \qquad \qquad \qquad \qquad  \textrm{$\mu$ non-integer}, \\
 \frac{{\rm sgn}(\mu)}{g_b^2}  \left( \frac{x_{|\mu|-1,1} \ell}{2z_0} \right)^{2|\mu|-2} \frac{2\pi Y_{|\mu|-1}(x_{|\mu|-1,1})}{|\mu| !(|\mu|-1)! x_{|\mu|-1,1} J'_{|\mu|-1}(x_{|\mu|-1,1})} \ \frac{m_1^2}{k^2-m_1^2}+\cdots~, \\
 \qquad \qquad \qquad \qquad \qquad \qquad \qquad \qquad \qquad \qquad \qquad \qquad \qquad  \textrm{$\mu$ integer},
 \end{array} \right.
\end{align}
where $J'$ is the derivative of the Bessel function with respect to its argument.  On comparing this with \eqref{MEpoles}, we learn that $\mu$ should be taken negative, and that the decay constant for the first resonances has the form
\begin{align}\label{decayconst}
f_{b}^{(1)}(m_1)=\left\{
\begin{array}{l}
\frac{1}{g_b}\left( \frac{\ell}{z_0}\right)^{|\mu|-1}\left[ \left(\frac{x_{|\mu|-1,1}}{2} \right)^{|\mu|-1}\sqrt{\frac{2x_{|\mu|-1,1}\Gamma(-|\mu|)J_{1-|\mu|}(x_{|\mu|-1,1})}{\Gamma(|\mu|) J_{|\mu|-1}'(x_{|\mu|-1,1})}} \ \right]z_0^{-1}~, \\
\qquad \qquad \qquad \qquad \qquad \qquad \qquad \qquad \qquad \qquad \qquad   \textrm{$\mu$ non-integer}, \\
\frac{1}{g_b}\left( \frac{\ell}{z_0}\right)^{|\mu|-1}\left[ \left(\frac{x_{|\mu|-1,1}}{2} \right)^{|\mu|-1}\sqrt{-\frac{2x_{|\mu|-1,1}\pi Y_{|\mu|-1}(x_{|\mu|-1,1})}{|\mu|! (|\mu|-1)! J_{|\mu|-1}'(x_{|\mu|-1,1})}} \ \right]z_0^{-1}~, \\
\qquad \qquad \qquad \qquad \qquad \qquad \qquad \qquad \qquad \qquad \qquad   \textrm{$\mu$ integer}.
\end{array}\right.
\end{align}
We emphasize that $f_{b}^{(1)}(k)$ depends on the momentum, and that the prediction made here is for the on-shell value of the first resonance. Note also that the sign of $\mu$ turns out to be
significant: we must choose one sign over the other to have real decay constants\footnote{The fact that $\mu$ turns out to be negative -- which
implies that we are fixing the value of an anti-self-dual source on the UV boundary -- has no physical meaning, however: if we had instead chosen $\bar{b}$ as our fundamental field and performed
the entire analysis above for $\bar{b}$ instead of $b$, we would have instead found that we had to fix a self-dual source on the UV boundary.}.

Based on the non-interacting model we have described so far, we were able to make concrete predictions for the masses and decay constants of the $\omega'/\rho'$ and the $h_1/b_1$
meson in terms of the parameters $\mu$, $g_b$, and $z_0/\ell$. Note that if we set $|\mu|=1$,  its value in perturbative QCD,
we would find that both the $\omega'/\rho'$ and the $h_1/b_1$ states
are degenerate with the $\omega/\rho$ mesons generated by the vector current, whose experimental mass is approximately half of the $\omega'/\rho'$ mass!

Conversely, we can use the measured value of the $h_1/b_1$ mass of $m=1230$ MeV to estimate
that $|\mu|$ runs to $\approx 1.82$ at this scale, with
\begin{align}
f_{b}^{(1)}(m_1) \approx \frac{1}{g_b}\left( \frac{\ell}{z_0}\right)^{0.82} 2092 \ \text{MeV}~.
\end{align}
While we could use this data to now fix the value of another
parameter, we see that our description of the tensor operator is
still incomplete. So far we have introduced $O^T$ with a
non-interacting Lagrangian, reading off the masses and decay
constants of the resonances it creates; these particles are totally
ignorant of chiral symmetry breaking, and furthermore have no decay
modes. In order to make better contact with QCD, lifting the
degeneracy between the $1^{+-}$ and $1^{--}$ states and introducing
interactions that allow the vector- and tensor-generated $\omega/\rho$
mesons to mix, we must add 5d interaction terms to the bulk
Lagrangian. We will now discuss how to identify these terms.

\section{Interactions: discrete symmetries and $N_c$ counting}

In order to classify the modes of $b_{MN}$ corresponding to various
meson states, we must carefully understand the holographic
equivalents of
discrete symmetries in QCD.
 Invariance of the gravity dual action under these symmetries will also
 constrain the  interactions we can legally include in $S_{\rm int}$.

Consider the pure gauge action, ignoring the Chern-Simons terms:
\beq S_{\rm gauge}= -\frac{1}{4 \ell g_5^2} \int d^5x \sqrt{g} {\rm tr}
\biggl( F_L^2 + F_R^2 \biggr) ~.\eeq
This sector of the theory
has five $Z_2$ symmetries. One of these is $T$, time reversal
invariance, which will not be needed in what follows. There is also
five-dimensional parity, which acts to reverse the sign of an odd number of
the spatial coordinates; we take its action to be
\beq
\label{pdef} P_5:  A^M(x^0,\vec x,z)  \rightarrow
(-1)^{I(M)}A^M(x^0, -\vec x, z)~, \eeq
where $I(M)$ is $0$ if $M=0,z$
and $1$ if $M=1,2,3$. In addition there are three $Z_2$ symmetries
which do not act on the coordinates:
\bea \tilde P: A_L &
\leftrightarrow& A_R~, \cr
 C_L:  A_L & \rightarrow& - A_L^* ~,\cr
 C_R: A_R & \rightarrow& - A_R^*~. \eea
Regarding the last two, we write the
Lie algebra of the gauge group as
\beq [T^a,T^b] = i f^{abc} T^c ~,
\eeq
with the $T^a$ Hermitian, normalized to $\tr (T^a T^b) =
\delta^{ab}/2$ and the structure constants $f^{abc}$ real. It is
then easy to check that $-(T^a)^*$ obey the same Lie algebra as the
$T^a$ so the transformation $T^a \rightarrow - (T^a)^*$ is an
automorphism of the Lie algebra. For $SU(N)$ this transformation
takes the generators in the ${\bf N}$ to those in the ${\bf \overline
N}$. For $U(1)$ it just takes $A_\mu \rightarrow -A_\mu$ which is
the usual action of $C$ on the vector potential, so it makes sense
to call such a transformation charge conjugation. The action $S_{\rm
gauge}$ has separate charge conjugation symmetries for $U(N_f)_L$
and $U(N_f)_R$.

We now consider adding the Chern-Simons term
\beq
S_{\textrm{CS}}= \frac{N_c}{24 \pi^2} \int_{\mathcal{M}} \biggl( \omega_5(A_L) - \omega_5(A_R) \biggr)~.
\eeq
This term changes sign under $P_5$ which changes the orientation of $\mathcal{M}$ and it also clearly changes
sign under $\tilde P$. Thus this term is only invariant under the combination $P \equiv P_5 \tilde P$. This is analogous to
the discussion in \cite{wzwterm} of $Z_2$ symmetries in the pion low-energy effective action where one finds an extra $Z_2$ symmetry of the
action which is not a symmetry of QCD, and which is only broken by the addition of the Wess-Zumino-Witten term.

Since $\tr F_L^3 = d \omega_5(A_L)$ and under $C_L$ we have $\tr
F_L^3 \rightarrow - \tr F_L^T F_L^T F_L^T = - \tr F_L^3 $ we see
that $\omega_5(A_L)$, $\omega_5(A_R)$ are odd under $C_L$, $C_R$.
Thus the Chern-Simons term is invariant under the combinations $C_L
C_R \tilde P$, or under $C_L C_R P$.  We will pick the first one and
call it $C$, $C \equiv C_L C_R \tilde P$. Thus the pure gauge
action, including the Chern-Simons term, is invariant under a $Z_2
\times Z_2$ symmetry generated by $C$ and $P$.

We now extend these symmetries to the bifundamental fields $X,b$.
The covariant derivative term $ |DX|^2$ is invariant provided that
$C: X \rightarrow \pm X^T, P: X \rightarrow \pm X^\dagger \equiv \pm \overline{X}$, but
since changing the sign of $X$ can be accomplished by a gauge
transformation we can choose the action to be
\bea C: & X \rightarrow  X^T~, \cr
P: & X \rightarrow  \overline{X} ~. \eea
where the action of $P$ on the argument of $X$ is the same as $P_5$ in
\eqref{pdef} and will be suppressed from now on. Note that if we
follow the treatment in \cite{hardwall} and expand $X$ around its
vacuum expectation value as $X= X_0(z) \exp(2 i \pi^a T^a)$ then the
above action of $C,P$ agrees with the canonical assignments of $C,P$
to the pion fields\footnote{Later on we will find it more convenient
to make a field redefinition which embeds the 4d pion field entirely
in the axial-vector gauge field.}. The terms in the action involving
$b$ are invariant under $C,P$ provided that
\bea C: & b_{MN} & \rightarrow \pm b_{MN} ^T~, \cr
P: & b_{MN} & \rightarrow \pm (-1)^{I(M,N)} \bar{b}_{MN}~. \eea
where $I(M,N)$ is the number of
$(i,j)$ indices in $(MN)$.  We can fix the sign in the action of $C$
by comparison to the standard assignment of $C=-1$ to the $h_1$
meson. The action of $P$ on the real and imaginary parts of $b_{MN}$
differs by a sign, so the choice of sign in the action of $P$ just
changes which fields we associate to the real and imaginary parts of
$b_{MN}$. In what follows we choose the plus sign so that
\bea C:  & b_{MN} & \rightarrow - b_{MN} ^T~, \cr
P:  & b_{MN}  & \rightarrow (-1)^{I(M,N)} \bar{b}_{MN}~. \eea

Any bottom-up 5d dual contains, in principle, an infinite number of
possible interaction terms. Such terms are not only relevant to
higher point functions between QCD operators: when $\bar{q}q$
acquires a vev, breaking chiral symmetry, these interactions also
contribute to two-point functions, modifying the location of the
poles and values of the decay constants. In the original hard wall
model \cite{hardwall}, this mechanism breaks the degeneracy between
vector and axial-vector modes. Once we include $b_{MN}$, such terms
will induce the mixing between vector meson states generated by
$O^T$, and those generated by $O^V$.

It is not trivial to decide which of this infinite set of terms to
include in our analysis. In the language of effective field theory
we often group interactions by dimension and assume that higher
dimension operators are suppressed by some scale. From the action in
\eqref{snought} we find that the mass dimensions of the fields are
\beq [X]=3/2~, \qquad [b]=3/2~, \qquad [F]=2~. \eeq
so in principle we
should only consider interaction terms, such as those roughly of the
form $bFX$, which have dimension 5.
No real separation of scales exists in the hard wall model, however,
so it is not clear that this classification of dominant interaction
terms is valid.

Another possible organizing principle is to consider instead the
large-$N_c$ scaling of interaction terms, and to include only those
leading in $1/N_c$.
We first recall the standard large $N_c$ counting rules for QCD as
reviewed in \cite{Coleman, Manohar}. Consider quark bilinear
operators, $\hat H_i$, which can include any number of gauge fields
as well. The large $N_c$ rules for QCD diagrams give
\beq \langle \hat H_1 \cdots \hat H_r \rangle \sim N_c^{1-r}~. \eeq
The operators
$\hat H_i$ are normalized so that $\sqrt{N_c} \hat H_i$ has an
amplitude of order one to create a one meson state. Therefore matrix
elements of operators that create mesons with unit amplitude behave
as
\beq \label{propnorm} \langle \sqrt{N_c} \hat H_1 \cdots
\sqrt{N_c} \hat H_r \rangle \sim N_c^{1-r/2}~. \eeq
The two-point function is $O(1)$ and the three-point function is
$O(1/\sqrt{N_c})$. This shows that the coupling is
$O(1/\sqrt{N_c})$, and a two-body decay amplitude for a meson is
$O(1/\sqrt{N_c})$ so mesons become stable in the large $N_c$ limit.

In the hard wall model we have been using a Lagrangian schematically
of the form
\beq S \sim \frac{1}{g_5^2} \int F^2 + \frac{1}{g_b^2}
\int b^2 + \int( |DX|^2 + |X|^2 )+ \cdots ~, \eeq
where $\cdots$
represents additional interaction terms. If we define rescaled gauge
fields $A'$ via  $A= g_5 A'$ and a rescaled tensor field via $b=g_b
b'$ then we have
\beq S \sim \int F'^2 + \int b'^2 + \int (|DX|^2 +
|X|^2)  + \cdots~. \eeq
By computing two- and three-point functions and matching to the
$N_c$ counting of the field theory, one finds $g_5 \sim
1/\sqrt{N_c}$ which is consistent with the result found in \cite{hardwall} from matching the UV
behavior of the vector current two-point function. The field $b$ only appears quadratically in what we have done so far, so the
$N_c$ dependence of  $g_b$ is not determined.
If we use the above Lagrangian to compute two-point functions
by the usual prescription in AdS/CFT they will be order $N_c^0$ in
agreement with \eqref{propnorm}. Three-point functions, which
involve couplings between three gauge fields, or a gauge field and
two pion fields and so on, will involve a factor of $g_5$ from the
rescaling, and so will indeed be of order $1/\sqrt{N_c}$.

Terms involving the tachyon field $X$ are more subtle.  Consider for example a
coupling of the form $b X F$ which can contribute to
two-point functions when $X$ is set equal to its vev, and apparently also
to three-point couplings when we include fluctuations of $X$ (pion
modes).  If we only focus on pion couplings then we can write $X=X_0e^{2 i \pi}$
which we leave invariant while redefining the gauge fields and $b$
field by what would be a gauge transformation if we were
transforming $X$ as well and with $V_L=V_R^\dagger=e^{i \pi}$. This
removes the pion field from $X$ and puts it entirely in the axial
gauge field. After this redefinition, we can replace any $X$
appearing in the action with $X_0$. Now the $b X F$ term
contributes only to the two-point function, and one might conclude that its coupling should be
of order $N_c^0$. However in general one could also look at fluctuations in the magnitude of $X$ which
are dual to the broad $\sigma$ resonance of QCD. This term then contributes to three-point couplings involving $ \sigma$, a vector meson and a $h_1/b_1$ meson and one
concludes that in fact its coupling should be of order $1/\sqrt{N_c}$. This term thus gives a subleading contribution
to the two-point function in the $1/N_c$ expansion. Terms containing even more powers of $X$ will be suppressed by
additional powers of $1/\sqrt{N_c}$ and should be small at large $N_c$.

We now consider two interaction terms
that are consistent with all global and local symmetries and are the leading terms that  lead
to tensor-vector mixing in the $\omega/\rho$ sector, and break the
degeneracy between the $h_1/b_1$ and tensor-$\omega/\rho$ spectrum.
While conceptually straightforward, a thorough analysis of the
interacting model turns out to be technically complex, and is
postponed to a forthcoming paper \cite{inprep}.

The first term we consider  is the  unique cubic, dimension five operator which is $P$, $C$ and $U(N_f)_L \times U(N_f)_R$ invariant:
\beq \label{Sgone}
S_{g_1} = g_1\int d^5x \sqrt{g} \textrm{tr}\left\{ b_{MN} F_{R}^{MN} \overline{X} + \overline{X} F_{L}^{MN} b_{MN} + X F_{R}^{MN} \bar{b}_{MN} + \bar{b}_{MN} F_{L}^{MN} X \right\}~.
\eeq
By the earlier argument after rescaling fields we find $g_1 g_b g_5 \sim O(N_c^{-1/2})$ which implies $g_1 g_b \sim O(N_c^0)$.
Evaluating $X$ on its chiral symmetry breaking vev,
$\langle X \rangle = \langle \overline{X} \rangle \propto {\bf 1}  v(z)$, one finds a quadratic order
term that mixes the vector gauge field $V$ with the two-form $b$.  The mixing modifies the mass spectrum
of both vector and tensor-$\omega/\rho$ states, and breaks the degeneracy between the spectrum of $h_1/b_1$
and tensor-$\omega/\rho$ states.

The second interaction term we consider is the dimension six term
\bea \label{Sgtwo}
S_{g_2} = g_2 \ell \int d^5 x \sqrt{g} \tr \left\{ b_{MN} \overline{X} b^{MN} \overline{X} + \bar{b}_{MN} X \bar{b}^{MN} X \right\}~,
\eea
and we have $g_2 g_b \sim O(N_c^{-1/2})$. 
Although this term does not contribute directly to tensor-vector mixing in the $\omega/\rho$ sector, it does contain quadratic terms, when $X$ is evaluated on its vev, that break the degeneracy between the $h_1/b_1$ and tensor-$\omega/\rho$ sectors.  

 In a follow-up paper we will analyze the interacting model
\eqref{snought}, with $S_{\rm int} = S_{g_1} + S_{g_2}$.

\section{Conclusions}

We have presented a consistent formalism for including fields dual
to the antisymmetric tensor quark bilinear $O^T$ in a
five-dimensional gravity dual description of QCD. This operator
creates $J^{PC}=1^{+-}$ mesons such as the $h_1$ and $b_1$ mesons as
well as $J^{PC}=1^{--}$ $\omega/\rho$-like mesons.

While the principle of including this new field in the holographic
framework is relatively straightforward, the implementation required
several novel strategies. One was the use of a first order action,
required for gauge invariance under
$U(N_f)_L \times U(N_f)_R$ flavor transformations while demanding
that the fields give rise to the correct number of degrees of
freedom in four dimensions. Another key technical point was to determine
a unique IR boundary condition for the new field, which was based on
the physical requirement that $O^T$ should not generate a
zero-momentum mode. We should also note that for $O^T$ we are forced
to account for the running of the anomalous dimension, rather than
simply fitting to the free field dimension as suffices for mesons
created by conserved currents.

We also performed an analysis of the interaction terms allowed by
the discrete and gauge symmetries of the model, and classified the
lowest dimension and leading order in $1/N_c$ terms that mix different states with the quantum
numbers of the $\rho, \omega$ mesons in the presence of chiral
symmetry breaking. In principle this model can now make new
predictions for a number of physical quantities including
\begin{itemize}
\item the mass spectrum and decay constants for the $h_1,b_1$ mesons,
\item the rates for $h_1 \rightarrow \rho + \pi$ and $b_1 \rightarrow \omega + \pi$ as
well as the $D/S$ amplitude ratio for these decays,
\item the mass spectrum and vector and tensor decay constants for the low-lying
excited $\rho$ and $\omega$ meson states.
\end{itemize}
Unfortunately this analysis requires rather complicated numerical analysis and as a
result we postpone presentation of these results to a later paper \cite{inprep}.

Finally, we reiterate that although we have focused on the hard wall model, our
considerations apply more generally
to a broad class of dual models of QCD. The soft wall model and other models that modify
the dilaton and/or metric require
antisymmetric tensor fields of the type described here in order to incorporate $h_1/b_1$ mesons.
It would also be interesting to
see whether the action described here for these fields can be derived for top-down models such as
the Sakai-Sugimoto
model using the techniques of open string field theory.

\bigskip
\bigskip
\noindent
{\bf Acknowledgements}

\noindent
JH would like to thank the LPTHE of the Universit\'{e} Marie et
Pierre Curie, and SKD Imperial College for hospitality during the
completion of this work.  AR acknowledges support from DOE grant 
DE-FG02-96ER50959.

\bigskip
\bigskip

\appendix

\section{Projection operators: definitions and useful relations}\label{ap:proj}

We will often find it useful to consider the (anti) self-dual or longitudinal/transverse pieces of the two-form $b_{\mu\nu}$. We can
define these in terms of the operators $P^{\pm}$ and $\mathcal{P}^{\parallel,\perp}$:

\begin{align}\label{eq:projectors}
&(\mathcal{P}^\parallel)_{\alpha\beta}^{\mu\nu}=2 k_{[\alpha} k^{[\mu} \delta_{\beta]}^{\nu]}~, \nonumber \\
&(\mathcal{P}^\perp)_{\alpha\beta}^{\mu\nu}= (k^2 {\bf 1} -\mathcal{P}^\parallel)_{\alpha\beta}^{\mu\nu}= k^2 \delta_{[\alpha}^\mu\delta_{\beta]}^\nu- 2 k_{[\alpha} k^{[\mu} \delta_{\beta]}^{\nu]}~, \nonumber \\
&(P^\pm)_{\alpha\beta}^{\mu\nu}=\frac{1}{2}\left( \delta_{[\alpha}^\mu\delta_{\beta]}^\nu\pm \frac{i}{2}\epsilon_{\alpha\beta}^{\phantom{\alpha\beta}\mu\nu}\right)~.
\end{align}
Useful relations among the projectors include
\begin{align}\label{eq:projrelns}
P^\pm \mathcal{P}^\perp P^\pm = \frac{k^2}{2} P^\pm~, \qquad P^\pm \mathcal{P}^\perp P^\mp = \frac{1}{2} (\mathcal{P}^\perp - \mathcal{P}^\parallel) P^\mp~.
\end{align}

The projectors $P^{\pm}$ are idempotent, $P^2 = P$, while the projectors $\mathcal{P}^{\parallel,\perp}$ have a
nonstandard normalization: $\mathcal{P}^2 = k^2 \mathcal{P}$.  The reason for this unaesthetic convention is that
we do not wish projectors to harbor massless poles.  To elucidate some of the potential subtleties, consider the
derivation of the expression for $\Pi^{T,T\perp}$ in \eqref{MEpoles}, starting with the definitions \eqref{Oodd},
\eqref{MECorRelation}, \eqref{ReducedME}.  By assumption, the states $|b_{1}^{(n),c}(k) \rangle$ form a complete
(properly normalized) basis for one-particle states that can be created by $O_{\mu\nu}^{T\perp,a}$ from the vacuum.
Therefore via insertion of the identity operator,
\begin{equation}\label{insertid}
\langle 0 | O_{\mu\nu}^{T\perp,a}(x) O_{\rho\sigma}^{T\perp,b}(0) |
0 \rangle = \delta^{ab} \sum_{n,\varepsilon} \int
\frac{d^3p}{(2\pi)^3} \frac{e^{-ip\cdot x}}{2E_n({\bf p})} (
f_{b}^{(n)} )^2 \ \frac{1}{2} \epsilon_{\mu\nu\alpha\beta}
\bar{\varepsilon}_{(b)}^{(n)\alpha} p^\beta \
\epsilon_{\rho\sigma\delta\gamma} \varepsilon_{(b)}^{(n)\delta}
p^\gamma~,
\end{equation}
where $E_{n}({\bf p}) = \sqrt{ {\bf p}^2 + (m_{b}^{(n)})^2 }$.  The
sum over (on-shell) polarizations yields
\begin{align}\label{polbsum}
\frac{1}{2} \sum_{\varepsilon} \epsilon_{\mu\nu\alpha\beta}
\bar{\varepsilon}_{(b)}^{(n)\alpha} p^\beta \
\epsilon_{\rho\sigma\delta\gamma} \varepsilon_{(b)}^{(n)\delta}
p^\gamma  =&~ (m_{(b)}^{(n)})^2 \eta_{\mu [\rho} \eta_{\sigma]\nu} -
( \eta_{\mu [\rho} p_{\sigma]} p_\nu - \eta_{\nu [\rho} p_{\sigma]}
p_\mu ) \cr \equiv &~
(\mathcal{P}^{\perp,(n)})_{\mu\nu,\rho\sigma}~.
\end{align}
Using Cauchy's theorem, we can convert \eqref{insertid} to an integral over four-momentum,
\begin{align}\label{corderstep2}
\langle 0 | O_{\mu\nu}^{T\perp,a}(x) O_{\rho\sigma}^{T\perp,b}(0) |
0 \rangle =&~ \delta^{ab} \sum_n \int \frac{d^4 p}{(2\pi)^4} \frac{i
e^{-ip\cdot x} (f_{(b)}^{(n)})^2 }{p^2 - (m_{(b)}^{(n)})^2}
(\mathcal{P}^{\perp,(n)})_{\mu\nu,\rho\sigma} \cr =&~ i \delta^{ab}
\int \frac{d^4 p}{(2\pi)^4} e^{-ip\cdot x}
(\mathcal{P}^{\perp})_{\mu\nu,\rho\sigma} \sum_n \frac{
(f_{(b)}^{(n)})^2 }{p^2 - (m_{(b)}^{(n)})^2 }~.
\end{align}
Plugging this into \eqref{MECorRelation} and using \eqref{ReducedME}, we straightforwardly recover the expression for $\Pi^{T,T \perp}$ in \eqref{MEpoles}.

In the second step of \eqref{corderstep2} we used the fact that $\oint dz f(a)/(z-a) = \oint dz f(z)/(z-a)$, which holds provided $f$ is a holomorphic function in the region bounded by the contour.  This allows us to take the projector for each mode off shell, so that it is a common factor which may be pulled out in front of the summand.  This step would have failed if $\mathcal{P}$ was a properly normalized projector, since then it would have a pole inside the contour of integration.

\section{General solution to the free equation of motion for $b_{MN}$} \label{ap:gensol}

The free equations of motion \eqref{beom} for the two-form are
\begin{align}
& \label{bmunueom} z\epsilon_{\mu\nu}^{\phantom{\mu\nu}\rho\sigma}\left( -2ik_\rho b_{\sigma z} +\dt_z b_{\rho\sigma}\right)-2i\mu b_{\mu\nu}=0~, \\
& \label{bmuzeom} z\epsilon_\mu^{\phantom{\mu}\nu\rho\sigma}k_\nu b_{\rho\sigma}+2\mu b_{\mu z} = 0~,
\end{align}
where we have used the Fourier-transformed fields
\begin{equation}
b_{MN}(k,z) =\int d^4x e^{ik \cdot x} b_{MN}(x,z)~.
\end{equation}

Equation (\ref{bmuzeom}) yields the constraint
\begin{align}\label{bmuz}
b_{\mu z}=-\frac{z}{2\mu } \epsilon_\mu^{\phantom{\mu}\nu\rho\sigma}k_\nu b_{\rho\sigma}~,
\end{align}
which, when plugged into the first equation of motion gives
\begin{align}\label{singleeom}
z^2\left(k_\mu k^\rho b_{\rho\nu}-k_\nu k^\rho b_{\rho\mu}-k^2 b_{\mu\nu}\right)+\frac{i\mu z}{2} \epsilon_{\mu\nu}^{\phantom{\mu\nu}\rho\sigma} \d_z b_{\rho\sigma}+\mu^2b_{\mu\nu}=0~,
\end{align}
or in terms of projectors,
\begin{align} \label{singleeom2}
& \left[ - z^2 \mathcal{P}^\perp + \mu z (P^+ - P^-) \d_z + \mu^2 \right] b_{\mu\nu} = 0~.
\end{align}

We can project \eqref{singleeom2} onto its self-dual and anti-self-dual parts.  For the projections of the $\mathcal{P}^\perp b$ term, we write $b = (P^+ + P^-)b$ and use \eqref{eq:projrelns} as necessary.  The resulting two equations may be rearranged to the form
\begin{align} \label{firstorder}
& b^{\mp} = \frac{2}{k^4 z^2} \left( \pm \mu z \d_z + \mu^2 - \frac{1}{2} k^2 z^2 \right) (\mathcal{P}^\perp - \mathcal{P}^\parallel) b^{\pm}~.
\end{align}
We can thus plug the equation for $b^-$ into the equation for $b^+$,
deriving an equation for $b^+$ alone, or vice versa.  After some
simplification and using $(\mathcal{P}^\perp - \mathcal{P}^\parallel)^2 = k^4 {\bf 1}$,  we arrive
at the equations derived in \cite{AFII} and quoted in
\eqref{bpmeoms}:
\begin{align} \label{secondorder}
& \left[ z^2 \d_{z}^2 - z \d_z + k^2 z^2 - \mu(\mu \pm 2) \right] b^{\pm}_{\mu\nu} = 0~.
\end{align}
The general solutions are
\begin{align} \label{generalbsols}
& b_{\mu\nu}^+(k,z) = \tilde{S}_{\mu\nu}(k) z J_{-\mu -1}(k z) + \tilde{s}_{\mu\nu}(k) z J_{\mu+1}(k z)~, \nonumber \\
& b_{\mu\nu}^{-}(k,z) =  \tilde{A}_{\mu\nu}(k) z J_{-\mu +1}(k z) + \tilde{a}_{\mu\nu}(k) z J_{\mu-1}(k z)~,
\end{align}
where $(\tilde{S},\tilde{s})$ and $(\tilde{A},\tilde{a})$ are self-dual and anti-self-dual polarizations respectively,
and $J$ is a standard Bessel function.  These solutions are appropriate (and convenient) for non-integer $\mu$.
For integer $\mu$, the $J_{-\mu \mp 1}$ terms should be replaced by $Y_{\mu \pm 1}$.  We must remember, however, that
we started with first order equations \eqref{firstorder}.  These equations will be satisfied on the
solutions \eqref{generalbsols} if and only if
\begin{align}
& (\tilde{A},\tilde{a}) = \frac{1}{k^2} (\mathcal{P}^\perp - \mathcal{P}^\parallel) (\tilde{S},\tilde{s}) \quad \Rightarrow \quad  (\tilde{S},\tilde{s}) = \frac{1}{k^2} (\mathcal{P}^\perp - \mathcal{P}^\parallel) (\tilde{A},\tilde{a})~.
\end{align}

Let us briefly comment on the bulk to boundary propagator for $b$.  The form it takes depends on the sign of $\mu$.  If $\mu > 0$, then the $\tilde{S}$ term of \eqref{generalbsols} dominates as $z = \varepsilon \to 0$.  The UV boundary condition matches $b^+$ onto a self-dual source, while the IR boundary condition sets $b^-$ to zero:
\begin{equation}\label{bbcsmuplus}
b_{\mu\nu}^+(k,\varepsilon) = \frac{\ell^{\mu-1/2}}{\varepsilon^\mu} S_{\mu\nu}(k)~, \qquad b_{\mu\nu}^-(k,z_0) = 0~, \qquad (\mu > 0)~.
\end{equation}
These conditions lead to the particular solution \eqref{btob}.  On the other hand, if $\mu <0$, the $\tilde{a}$ term dominates.  In this case $b^-$ should match onto an anti-self-dual source at the UV boundary and $b^+$ should be zero at the IR boundary:
\begin{equation}\label{bbcsmuminus}
b_{\mu\nu}^-(k,\varepsilon) = \frac{\ell^{|\mu|-1/2}}{\varepsilon^{|\mu|}} a_{\mu\nu}(k)~, \qquad b_{\mu\nu}^+(k,z_0) = 0~, \qquad (\mu < 0)~.
\end{equation}
These boundary conditions lead to the particular solution
\begin{align}\label{btobmuminus}
& b_{\mu\nu}^-(k,z) =   \ell^{|\mu| -1/2} a_{\mu\nu}(k) \frac{B_{<}^-(k,z)}{\varepsilon^{|\mu|} B_{<}^-(k,\varepsilon)}~, \nonumber \\
& b_{\mu\nu}^+(k,z) = \ell^{|\mu| -1/2} \left[ a_{\mu\nu} - \frac{2}{k^2} \left( k_\mu k^\rho a_{\rho\nu} - k_\nu k^\rho a_{\rho\mu} \right) \right] \frac{B_{<}^+(k,z)}{\varepsilon^{|\mu|} B_{<}^-(k,\varepsilon)}~,
\end{align}
where
\begin{align}\label{Blessdef}
& B_{<}^{\pm}(k,z) = z J_{-|\mu| \pm 1}(k z) - c_{b}^{<}(k,z_0) z J_{|\mu| \mp 1}(k z)~, \qquad {\rm with} \nonumber \\
& c_{b}^{<}(k,z_0) = \frac{ J_{-|\mu|+1}(k z_0) }{J_{|\mu|-1}(k z_0) }~.
\end{align}
These expressions are appropriate for the non-integer $\mu$ case.  For integer $\mu$, the $J_{-|\mu| \pm 1}$ should be replaced by $Y_{|\mu| \mp 1}$.

\section{Dual sources, correlation functions, and matrix elements}\label{ap:MEcomp}

As usual in AdS/CFT, we determine the generating functional for the
correlators of field theory operators by evaluating the supergravity
action on the bulk-to-boundary propagators of the dual fields. The
UV boundary conditions on these propagators are determined in such a
way that for a 4d source $\phi_0(x)$, the dual five-dimensional
field has boundary condition\footnote{Up to powers of $\varepsilon$, as discussed after equation \eqref{btob}.} $\phi (x,z=\varepsilon)=\phi_0(x)$.
This determines the generating functional to be
\begin{align}
Z[\phi_0(x)]=e^{i\left.S_{sugra} (\phi)\right|_{\phi(\varepsilon)=\phi_0}}~.
\end{align}
(Strictly speaking, this would be the leading saddle point approximation in a full quantum gravity dual description).
We will usually work with momentum-space correlators. Functional differentiation in terms of momentum-space sources obeys
\begin{align}
\frac{\delta}{\delta J(-k)} J(p) = (2\pi)^4 \delta^{(4)}(p-k)~.
\end{align}

We are mostly interested in correlators of the tensor operator
$\bar{q}\sigma_{\mu\nu}q$, but we often work with the pseudo-tensor
or (anti) self-dual versions discussed in Section 2. Here we review
the definitions of the sources for these operators and the relations
between them.

In the case $\mu >0$ the bulk to boundary propagator is determined by a self-dual source, $P^+S=S$, which naturally couples to a self-dual operator in the field theory Lagrangian (in order to get a Lorentz scalar).  The conjugate $\bar{S}_{\mu\nu}$ is anti-self-dual and naturally couples to the anti-self-dual, conjugate operator.  These operators can be taken as
\begin{equation}\label{Opm}
O^{\pm}_{\mu\nu}(x) = \bar{q}(x) \sigma_{\mu\nu} \frac{( 1 \pm \gamma_5  )}{2} q(x)~,
\end{equation}
and one indeed has $(O^{T,+})^\ast = O^{T,-}$.
From them one can construct tensor and pseudo-tensor operators,
\begin{equation}\label{OTOPT}
O_{\mu\nu}^T = O_{\mu\nu}^+ + O_{\mu\nu}^- = \bar{q}(x) \sigma_{\mu\nu} q(x)~, \qquad O_{\mu\nu}^{PT} = O_{\mu\nu}^+ - O_{\mu\nu}^- = \bar{q}(x) \sigma_{\mu\nu} \gamma_5 q(x)~.
\end{equation}
These should couple to a tensor source and pseudo-tensor source, $\mathcal{T}_{\mu\nu}$ and $\mathcal{P}_{\mu\nu}$, given by
\begin{equation}\label{TPTtoSSbar}
\begin{array}{l} \mathcal{T}_{\mu\nu}(k) = \frac{1}{2} \left( S_{\mu\nu}(k) + \bar{S}_{\mu\nu}(-k) \right) \\
\mathcal{P}_{\mu\nu}(k) = \frac{1}{2} \left( S_{\mu\nu}(k) - \bar{S}_{\mu\nu}(-k) \right) \end{array} ~~ \Rightarrow ~~
\begin{array}{l} S_{\mu\nu}(k) = \mathcal{T}_{\mu\nu}(k) + \mathcal{P}_{\mu\nu}(k) \\
\bar{S}_{\mu\nu}(k) =  \mathcal{T}_{\mu\nu}(-k) - \mathcal{P}_{\mu\nu}(-k) \end{array} ~, \quad (\mu > 0)~.
\end{equation}
Note that $\mathcal{P},\mathcal{T}$ satisfy $\mathcal{P}_{\mu\nu} = \frac{i}{2} \epsilon_{\mu\nu}^{\phantom{\mu\nu}\rho\sigma} \mathcal{T}_{\rho\sigma}$, as do the corresponding operators, $O^{PT}$ and $O^T$.  The relative normalization between $\mathcal{T},\mathcal{P}$ and $S,\bar{S}$ is fixed by the requirement that
\begin{equation}\label{sourcenormfix}
S^{\mu\nu} O_{\mu\nu}^+ + \bar{S}^{\mu\nu} O_{\mu\nu}^- = \mathcal{T}^{\mu\nu} O_{\mu\nu}^T + \mathcal{P}^{\mu\nu} O_{\mu\nu}^{PT}~.
\end{equation}

When $\mu <0$ on the other hand, the bulk to boundary propagator is given in terms of an anti-self-dual source $a_{\mu\nu}$.  This source naturally couples to $O^{T,-}$, while $\bar{a}_{\mu\nu}$ couples to $O^{T,+}$.  The relationship of these sources to the tensor and pseudo-tensor sources introduced above is
\begin{equation}\label{TPTtoaabar}
\begin{array}{l} \mathcal{T}_{\mu\nu}(k) = \frac{1}{2} \left( a_{\mu\nu}(k) + \bar{a}_{\mu\nu}(-k) \right) \\
\mathcal{P}_{\mu\nu}(k) = \frac{1}{2} \left( \bar{a}_{\mu\nu}(k) - a_{\mu\nu}(-k) \right) \end{array} ~~\Rightarrow ~~
\begin{array}{l} a_{\mu\nu}(k) = \mathcal{T}_{\mu\nu}(-k) - \mathcal{P}_{\mu\nu}(-k) \\
\bar{a}_{\mu\nu}(k) =  \mathcal{T}_{\mu\nu}(k) + \mathcal{P}_{\mu\nu}(k) \end{array} ~, \quad (\mu < 0)~.
\end{equation}
(Note that \eqref{TPTtoSSbar} and \eqref{TPTtoaabar} \emph{do not} imply $a = \bar{S}$; these equations apply for different values of $\mu$, corresponding to different dual Lagrangians).

We now derive the tensor-tensor two-point function from the free
supergravity action.  The same result was already derived in \cite{AFII} , but we repeat it here for completeness.

We begin by determining the generating functional for the two-point functions.  First assume that $\mu >0$.  Evaluating the action on the bulk-to-boundary propagator \eqref{btob}, the bulk and IR boundary contributions to the action vanish, and we are left with the UV boundary term,
\begin{equation}\label{AFboundaryterm}
S_{\rm sd} = \frac{1}{4 g_{b}^2 \ell} \int d^4 x \tr
\left[ \bar{b}_{\mu\nu} b^{\mu\nu} \right]_{z = \varepsilon} = \frac{1}{4\ell g_{b}^2} \int d^4 x  \tr \left[
\overline{b_{\mu\nu}^+} b^{-\mu\nu} + \overline{b_{\mu\nu}^-} b^{+ \mu\nu}
\right]_{z = \varepsilon}  ~.
\end{equation}
Plugging in \eqref{btob}, evaluated at $z
= \varepsilon$ gives
\begin{align}\label{onshellbndry}
S_{\rm sd} =&~ \frac{\ell^{2\mu-2}}{4 g_{b}^2} \int \frac{d^4 k}{(2\pi)^4} \frac{B_{>}^-(k, \varepsilon)}{\varepsilon^{2\mu} B_{>}^+(k, \varepsilon)} \tr \displaystyle\biggl\{  \bar{S}^{\alpha\beta} \left[ S_{\alpha\beta} - \frac{2}{k^2} \left( k_\alpha k^\gamma S_{\gamma\beta} - k_\beta k^\gamma S_{\gamma\alpha} \right) \right] + c.c. \displaystyle\biggr\} \nonumber \\
=&~ - \frac{2 \ell^{2\mu-2}}{g_{b}^2} \int \frac{d^4 k}{(2\pi)^4}
\frac{B_{>}^-(k, \varepsilon)}{\varepsilon^{2\mu} B_{>}^+(k, \varepsilon)} \tr \left[ \frac{1}{k^2}
k_\nu \bar{S}^{\nu\mu} k^\rho S_{\rho\mu} \right]~,
\end{align}
where we have used the fact that a self-dual tensor contracted with an ant-self-dual one vanishes.
Since we are interested in the tensor-tensor two-point function, we
choose to express the generating functional in terms of the tensor
source.  Using \eqref{TPTtoSSbar} with the constraint $\mathcal{P}_{\mu\nu} = \frac{i}{2} \epsilon_{\mu\nu}^{\phantom{\mu\nu}\rho\sigma} \mathcal{T}_{\rho\sigma}$, a short calculation shows that
\begin{align}\label{Tplugin}
k_{\nu} \bar{S}^{\nu\mu} k^\rho S_{\rho\mu} =&~ - \frac{1}{2} \mathcal{T}^{\alpha\beta}(-k) \left[ k^2
\mathcal{T}_{\alpha\beta}(k) - 2 \left( k_\alpha
k^\gamma \mathcal{T}_{\gamma\beta}(k) - k_\beta k^\gamma
\mathcal{T}_{\gamma\alpha}(k) \right)  \right]~.
\end{align}
Plugging this into \eqref{onshellbndry}, we arrive at \eqref{Ssdonshell}:
\begin{equation}\label{STT}
S_{\rm sd} = \frac{\ell^{2\mu-2}}{g_{b}^2} \int  \frac{d^4
k}{(2\pi)^4} \frac{B_{>}^-(k, \varepsilon)}{k^2 \varepsilon^{2\mu} B_{>}^+(k,
\varepsilon)} \tr \left\{ \mathcal{T}^{\alpha\beta}(-k) \left[ \mathcal{P}_{\alpha\beta,\delta\gamma}^\perp -\mathcal{P}_{\alpha\beta,\delta\gamma}^\parallel  \right]
\mathcal{T}^{\delta\gamma}(k) \right\}  ~.
\end{equation}

Now suppose that $\mu <0$.  In this case the on-shell action is
\begin{equation}\label{AFboundaryterm2}
S_{\rm sd} = - \frac{1}{4\ell g_{b}^2} \int d^4 x  \tr \left[
\overline{b_{\mu\nu}^+} b^{-\mu\nu} + \overline{b_{\mu\nu}^-} b^{+ \mu\nu}
\right]_{z = \varepsilon}  ~.
\end{equation}
Plugging in the bulk to boundary propagator \eqref{btobmuminus}, we find
\begin{align}\label{onshellbndry2}
S_{\rm sd} =&~ - \frac{\ell^{2|\mu|-2}}{4 g_{b}^2} \int \frac{d^4 k}{(2\pi)^4} \frac{B_{<}^+(k, \varepsilon)}{\varepsilon^{2|\mu|} B_{<}^-(k, \varepsilon)} \tr \displaystyle\biggl\{  \bar{a}^{\alpha\beta} \left[ a_{\alpha\beta} - \frac{2}{k^2} \left( k_\alpha k^\gamma a_{\gamma\beta} - k_\beta k^\gamma a_{\gamma\alpha} \right) \right] + c.c. \displaystyle\biggr\} \nonumber \\
=&~  \frac{2 \ell^{2|\mu|-2}}{g_{b}^2} \int \frac{d^4 k}{(2\pi)^4}
\frac{B_{<}^+(k, \varepsilon)}{\varepsilon^{2|\mu|} B_{<}^-(k, \varepsilon)} \tr \left[ \frac{1}{k^2}
k_\nu \bar{a}^{\nu\mu} k^\rho a_{\rho\mu} \right]~,
\end{align}
Using \eqref{TPTtoaabar} with  $\mathcal{P}_{\mu\nu} = \frac{i}{2} \epsilon_{\mu\nu}^{\phantom{\mu\nu}\rho\sigma} \mathcal{T}_{\rho\sigma}$,
\begin{align}\label{Tplugin2}
k_{\nu} \bar{a}^{\nu\mu} k^\rho a_{\rho\mu} =&~ - \frac{1}{2} \mathcal{T}^{\alpha\beta}(k) \left[ k^2
\mathcal{T}_{\alpha\beta}(-k) -2\left( k_\alpha
k^\gamma \mathcal{T}_{\gamma\beta}(-k) - k_\beta k^\gamma
\mathcal{T}_{\gamma\alpha}(-k) \right)  \right]~,
\end{align}
and thus,
\begin{equation}\label{STT2}
S_{\rm sd} = - \frac{\ell^{2|\mu|-2}}{g_{b}^2} \int  \frac{d^4
k}{(2\pi)^4} \frac{B_{<}^+(k, \varepsilon)}{ k^2 \varepsilon^{2|\mu|} B_{<}^-(k,
\varepsilon)} \tr \left\{ \mathcal{T}^{\alpha\beta}(-k) \left[ \mathcal{P}_{\alpha\beta,\delta\gamma}^\perp - \mathcal{P}_{\alpha\beta,\delta\gamma}^\parallel  \right]
\mathcal{T}^{\delta\gamma}(k) \right\}  ~.
\end{equation}

The two cases $\mu >0$, $\mu <0$ can be combined as follows.  Define the functions
\begin{align}\label{BnBd}
& B_n(k,z) = \left\{ \begin{array}{l l} z J_{-|\mu| +1}(k z) - c_b(k, z_0) z J_{|\mu| - 1}(k z)~, & ~~ \mu~\textrm{non-integer,} \\  z Y_{|\mu|-1}(k z) - c_b(k, z_0) z J_{|\mu| - 1}(k z)~, & ~~ \mu~\textrm{integer,} \end{array} \right.  \nonumber \\
& B_d(k,z) = \left\{ \begin{array}{l l}  z J_{-|\mu|-1}(k z) - c_b(k,z_0) z J_{|\mu|+1}(k z)~, & ~~\mu~\textrm{non-integer,} \\  z Y_{|\mu|+1}(k z) - c_b(k,z_0) z J_{|\mu|+1}(k z)~, & ~~\mu~\textrm{integer,} \end{array} \right.
\end{align}
with
\begin{equation}\label{cbdeffinal}
c_b(k,z_0) = \left\{ \begin{array}{l l} \frac{ J_{-|\mu| + 1}(k z_0) }{ J_{|\mu| -1}(k z_0) }~, & ~~ \mu~\textrm{non-integer,} \\ \\
\frac{ Y_{|\mu| - 1}(k z_0) }{ J_{|\mu| -1}(k z_0) }~, & ~~ \mu~\textrm{integer,} \end{array} \right.
\end{equation}
for both positive and negative $\mu$.  Observe that when $\mu > 0$, $B_n = B_{>}^-$, and when $\mu < 0$, $B_n = B_{<}^+$.  Similarly, when $\mu >0$, $B_d = B_{>}^+$ and when $\mu < 0$, $B_d = B_{<}^-$.  Thus, \eqref{STT} and \eqref{STT2} may be summarized as
\begin{equation}\label{STTfinal}
S_{\rm sd} = {\rm sgn}(\mu) \frac{\ell^{2|\mu|-2} }{g_{b}^2} \int  \frac{d^4
k}{(2\pi)^4} \frac{B_n(k, \varepsilon)}{ k^2 \varepsilon^{2|\mu|} B_d(k,
\varepsilon)} \tr \left\{ \mathcal{T}^{\alpha\beta}(-k) \left[ \mathcal{P}_{\alpha\beta,\delta\gamma}^\perp - \mathcal{P}_{\alpha\beta,\delta\gamma}^\parallel  \right]
\mathcal{T}^{\delta\gamma}(k) \right\} ,
\end{equation}
valid for all $\mu$.  This is the expression we can functionally differentiate to compute
the $O^T$-$O^T$ two-point correlator.  Due to
the relation between $\mathcal{P},\mathcal{T}$ discussed above, the
$\langle O^{PT} O^{PT} \rangle$ and $\langle
O^{T} O^{PT} \rangle$ correlators follow from
the tensor-tensor correlator.

The two-point correlator is computed via the usual prescription:
\begin{equation}\label{OTtwopointgen}
 \langle O_{\mu\nu}^T(p) O_{\rho\sigma}^T(-q) \rangle = \frac{\delta}{\delta \mathcal{T}^{\mu\nu}(-p)} \frac{\delta}{\delta \mathcal{T}^{\rho\sigma}(q)} i S_{\rm sd}~.
\end{equation}
Evaluating the right-hand side, restoring isospin indices, leads to
\begin{align}\label{OTtwopoint}
\langle O_{\mu\nu}^{T,a}(p) O_{\rho\sigma}^{T,b}(-q) \rangle = i (2\pi)^4 \delta^{(4)}(p-q) \delta^{ab} {\rm sgn}(\mu) \frac{4 \ell^{2|\mu|-2} B_n(q, \varepsilon)}{g_{b}^2 q^2 \varepsilon^{2|\mu|} B_d(q,
\varepsilon)} \left( \mathcal{P}_{\mu\nu,\rho\sigma}^\perp - \mathcal{P}_{\mu\nu,\rho\sigma}^\parallel \right)~,
\end{align}
which gives a matrix element
\begin{equation}\label{MEprediction}
\Pi^{T,T,ab}_{\mu\nu,\rho\sigma}(k) = \delta^{ab} {\rm sgn}(\mu) \frac{4
\ell^{2|\mu|-2}B_n(k, \varepsilon)}{g_{b}^2 k^2 \varepsilon^{2|\mu|}  B_d(k,
\varepsilon)} \left[  \mathcal{P}_{\mu\nu,\rho\sigma}^\perp - \mathcal{P}_{\mu\nu,\rho\sigma}^\parallel \right]~.
\end{equation}
Given \eqref{ReducedME}, we have $\Pi^{T,T \parallel} = - \Pi^{T,T \perp}$, with
\begin{equation}\label{PiTTresult}
\Pi^{T,T \perp}(k) = {\rm sgn}(\mu) \frac{4 \ell^{2|\mu|-2} B_n(k, \varepsilon)}{g_{b}^2 k^2
\varepsilon^{2|\mu|} B_d(k, \varepsilon)}~.
\end{equation}

Now, finally, we take the $\varepsilon \rightarrow 0$ limit.  We
will first consider the case of non-integer $\mu$, so that
\begin{equation}\label{Brationimu}
\frac{B_n(k,\varepsilon)}{\varepsilon^{2|\mu|} B_d(k,\varepsilon) } =
\frac{J_{-|\mu| + 1}(k \varepsilon) - c_b(k,z_0) J_{|\mu| -1}(k
\varepsilon)}{\varepsilon^{2|\mu|} \left[ J_{-|\mu| -1}(k\varepsilon) -
c_b(k,z_0) J_{|\mu|+1}(k \varepsilon) \right]  } ~,
\end{equation}
For convenience, define $B_{n,d}(k,z) = z \tilde{B}_{n,d}(k,z)$.  We have that $B_n/B_d = \tilde{B}_n/\tilde{B}_d$.  Using series expansions of the Bessel functions, we find that the
leading $\varepsilon$ behavior in the denominator is
\begin{equation}\label{BratioDenni}
\varepsilon^{2|\mu|} \tilde{B}_d(k,\varepsilon) = \frac{2^{|\mu|
+1}}{\Gamma(-|\mu|)} k^{-|\mu|-1} \varepsilon^{|\mu| -1} \left( 1 +
\mathcal{O}(\varepsilon^2) \right)~.
\end{equation}
Meanwhile in the numerator we have
\begin{align}\label{BratioNumni}
\tilde{B}_n(k,\varepsilon) =&~ \frac{2^{|\mu| -1}}{\Gamma(-|\mu| +2)} k^{-|\mu| +1}
\varepsilon^{-|\mu| +1} \left(1 + \mathcal{O}(\varepsilon^2) \right) + \nonumber \\
&~ \qquad \qquad \qquad \qquad - \frac{c_b(k,z_0) }{2^{|\mu| -1} \Gamma(|\mu|)} k^{|\mu| -1} \varepsilon^{|\mu| -1}
\left( 1 + \mathcal{O}(\varepsilon^2) \right)~.
\end{align}
Hence the ratio takes the form
\begin{align}\label{Brationimu2}
\frac{B_n(k,\varepsilon)}{\varepsilon^{2|\mu|} B_d(k,\varepsilon) } =&~
\frac{\Gamma(-|\mu|)}{4 \Gamma(-|\mu|+2)} k^2 \varepsilon^{-2(|\mu|+1)}
\left(1 + \mathcal{O}(\varepsilon^2) \right) + \nonumber \\
&~ \qquad \qquad \qquad \qquad - c_b(k,z_0) \frac{ \Gamma(-|\mu|)}{2^{2|\mu|} \Gamma(|\mu|)} k^{2|\mu|} \left( 1 +\mathcal{O}(\varepsilon^2) \right)~.
\end{align}
Observe that all of the terms in the first series go as
$\varepsilon$ to a negative or positive power; there are no terms
that go as $\varepsilon^0$ for non-integer $\mu$.  The terms that
diverge with $\varepsilon$ represent contact terms which we are not
interested in.  Hence, the first series can be ignored in the
$\varepsilon \rightarrow 0$ limit.  Meanwhile, only the first term
of the second series survives the limit.  Hence,
\begin{equation}\label{Brationimu3}
\lim_{\varepsilon \rightarrow 0}
\frac{B_n(k,\varepsilon)}{\varepsilon^{2|\mu|} B_d(k,\varepsilon) } =
~ \textrm{contact terms} ~ - \frac{\Gamma(-|\mu|)}{2^{2|\mu|}
\Gamma(|\mu|)} k^{2|\mu|} c_b(k,z_0) ~.
\end{equation}

Next consider the integer $\mu$ case.  We have
\begin{equation}\label{Bratioimu}
\frac{B_n(k,\varepsilon)}{\varepsilon^{2|\mu|} B_d(k,\varepsilon) } =
\frac{Y_{|\mu| - 1}(k \varepsilon) - c_b(k,z_0) J_{|\mu| -1}(k
\varepsilon)}{\varepsilon^{2|\mu|} \left[ Y_{|\mu| +1}(k\varepsilon) -
c_b(k,z_0) J_{|\mu|+1}(k \varepsilon) \right]  }
\end{equation}
For integer $n$, the Bessel function $Y_n$ is defined by the series
\begin{align}\label{Yimu}
Y_n(x) =&~ - \frac{1}{\pi} \left( \frac{x}{2} \right)^{-n} \sum_{j=0}^{n-1} \frac{(n-j-1)!}{j!} \left( \frac{x}{2} \right)^{2j} + \frac{2}{\pi} \log{\left( \frac{x}{2} \right)} J_n(x) + \nonumber \\
& \qquad - \frac{1}{\pi} \left( \frac{x}{2} \right)^n
\sum_{j=0}^{\infty} \frac{ \left[ \psi_0(j+1) + \psi_0(n+j+1)
\right] }{j! (n+j)!} \left( \frac{-x}{2} \right)^{2j}~,
\end{align}
where $\psi_0$ is the digamma function.  For integer values,
$\psi_0(n) = -\gamma + H_{n-1} = -\gamma + \sum_{j=1}^{n-1}
\frac{1}{j}$, where $\gamma$ is the Euler-Mascheroni constant, and
$H_n$ a harmonic number.  We note that the first series begins at
order $x^{-n}$ and terminates at order $x^{n-1}$.  The $J_n$ term
and the second series begin at order $x^n$.  In the denominator
then, the leading divergence is
\begin{equation}\label{BratioDeni}
\varepsilon^{2|\mu|} \tilde{B}_d(k,\varepsilon) = - \frac{2^{|\mu| +1}
|\mu|!}{\pi} k^{-|\mu|-1} \varepsilon^{|\mu| -1} \left( 1 +
\mathcal{O}(\varepsilon^2) \right)~.
\end{equation}
In the numerator, all terms in the first series of $Y_{|\mu|-1}$ range
from $\varepsilon^{-|\mu| +1}$ to $\varepsilon^{|\mu| -2}$, and hence
represent contact terms.  The terms we are interested in, given \eqref{BratioDeni}, go as $\varepsilon^{|\mu|-1}$.  They are the leading terms in $J_{|\mu| -1}$,
(both the explicit $J_{|\mu|-1}$ in \eqref{Bratioimu} and the one
contained in \eqref{Yimu}), as well as the first term in the series on the second line of \eqref{Yimu}:
\begin{align}\label{BratioNumi}
\tilde{B}_n(k,\varepsilon) =&~ \cdots  + \left[ \frac{2}{\pi} \log{\left( \frac{k \varepsilon}{2} \right)} -
c_b(k,z_0) \right] \frac{1}{2^{|\mu| -1} (|\mu|-1)!} k^{|\mu| -1} \varepsilon^{|\mu| -1} \left( 1 + \mathcal{O}(\varepsilon^2) \right) + \nonumber \\
& \qquad  - \frac{( H_{|\mu|-1} -2 \gamma) }{\pi 2^{|\mu| -1} (|\mu|-1)!}
k^{|\mu| -1} \varepsilon^{|\mu| -1} \left( 1 +
\mathcal{O}(\varepsilon^2) \right)~,
\end{align}
where the $\cdots$ represent the contact terms.  Thus we have that
\begin{align}\label{Bratioimu2}
\lim_{\varepsilon \rightarrow 0}
\frac{B_n(k,\varepsilon)}{\varepsilon^{2|\mu|} B_d(k,\varepsilon) } =&~
 \textrm{contact terms} ~ + \nonumber \\
& - \frac{\pi}{2^{2|\mu|} |\mu|! (|\mu| -1)!}
\left[ \frac{2}{\pi} \log{\left( \frac{k \varepsilon}{2} \right)} -
c_b(k,z_0) - \frac{1}{\pi} (H_{|\mu| -1} -2\gamma) \right] k^{2|\mu|}~.
\end{align}
In a standard renormalization scheme with renormalization scale $M_r = \ell^{-1}$, this becomes
\begin{equation}\label{Bratioimu3}
\lim_{\varepsilon \rightarrow 0}
\frac{B_n(k,\varepsilon)}{\varepsilon^{2|\mu|} B_d(k,\varepsilon) } =
~ \textrm{contact terms} ~ + \frac{1}{2^{2|\mu|} |\mu|! (|\mu|-1)!} \left[
 \pi c_b(k,z_0) - \log{(k^2 \ell^2)}  \right]
k^{2|\mu|}~.
\end{equation}

In summary, (dropping the contact terms), we have
\begin{equation}\label{PiTTlim}
\Pi^{T,T\perp}(k) =  \left\{ \begin{array}{l l} - \frac{{\rm sgn}(\mu) \Gamma(-|\mu|)}{2^{2|\mu|-2} g_{b}^2 \Gamma(|\mu|)}
c_b(k,z_0)  (k \ell)^{2|\mu|-2}~, & \mu ~ \textrm{non-integer,} \\
& \\ \frac{{\rm sgn}(\mu)}{ 2^{2|\mu| -2} g_{b}^2 |\mu|! (|\mu|-1)!} \left[  \pi c_b(k,z_0) -
\log{(k^2 \ell^2)} \right]
(k\ell)^{2|\mu|-2}~, & \mu ~ \textrm{integer.} \end{array} \right.
\end{equation}
This reproduces \eqref{PiTTcb} for $\mu > 0$.  Plugging in \eqref{cbdeffinal}, we recover \eqref{PiTTfinalgenmu}.

\section{Spectrum of normalizable modes}
\label{ap:normmodes}

In this case we are interested in the normalizable part of the
solution, \eqref{generalbsols}.  In the following we will consider the $\mu >0$ case only; an analogous treatment can be performed for the $\mu <0$ case.  For $\mu > 0$, we restrict to the normalizable component of \eqref{generalbsols} by setting $S = A = 0$, resulting in
\begin{equation}\label{bpmnormalizable}
b_{\mu\nu}^+(k,z) = \tilde{s}_{\mu\nu}(k) z J_{\mu + 1}(m z)~, \qquad
b_{\mu\nu}^-(k,z) = \left[ \tilde{s}_{\mu\nu} + \frac{4}{k^2} k_{[\mu}
\tilde{s}_{\nu] \rho} k^\rho \right] z J_{\mu -1}(m z)~.
\end{equation}
Since $b_{\mu\nu} = b_{\mu\nu}^+ +
b_{\mu\nu}^-$ and $b_{\mu z} = - \frac{z}{2\mu}
\epsilon_{\mu}^{\phantom{\mu}\nu\rho\sigma} k_\nu b_{\rho\sigma}$,
this implies
\begin{align}\label{bmunubmuz}
b_{\mu\nu}(k,z) =&~ \frac{2\mu}{m} \tilde{s}_{\mu\nu} J_\mu(m z) + \frac{4}{k^2} k_{[\mu} \tilde{s}_{\nu]\rho} k^\rho z J_{\mu -1}(m z)~, \\
b_{\mu z}(k,z) =&~ - \frac{1}{m}
\epsilon_{\mu}^{\phantom{\mu}\nu\rho\sigma} k_\nu \tilde{s}_{\rho\sigma} z
J_{\mu}(m z) =  \frac{2 i}{m} \tilde{s}_{\mu\nu} k^\nu z J_{\mu}(m z)~.
\end{align}
These solutions solve \eqref{singleeom2}-\eqref{secondorder} with eigenvalue $k^2 \to m^2$.  The IR Neumann-like boundary condition, $b^-(z_0) = 0$, fixes the eigenvalues to be $m_{n} = x_{\mu-1,n}/z_0$.  For each eigenvalue, there is a corresponding self-dual polarization $\tilde{s}_{\mu\nu}^{(n)}(k)$, which contains six real degrees of freedom.

We would like to evaluate (the quadratic part of) \eqref{firstorderaction} on a sum over the eigenmodes.  The boundary terms vanish since we have restricted to normalizable modes and imposed the IR boundary condition.  To evaluate the bulk part of the action, we require the components of $db$ and $\star b$, evaluated on the solution \eqref{bmunubmuz}.  Working in momentum space we have
\begin{align}\label{dbcomps}
(db)_{\mu\nu\rho} =&~ 3 \d_{[\mu} b_{\nu\rho]} \rightarrow -i (k_{\mu} b_{\nu\rho} + k_{\nu} b_{\rho \mu} + k_\rho b_{\mu\nu} )  \nonumber \\
=&~  - \frac{2 i \mu}{m} (k_{\mu} \tilde{s}_{\nu\rho} + k_{\nu} \tilde{s}_{\rho \mu} + k_\rho \tilde{s}_{\mu\nu} ) J_{\mu}(m z)~, \\
(db)_{\mu\nu z} \rightarrow &~ -i (k_\mu b_{\nu z} - k_\nu b_{\mu z} ) + \d_z b_{\mu\nu} \nonumber \\
=&~  \frac{4}{m} k_{[\mu} \tilde{s}_{\nu]\rho} k^\rho z J_{\mu}(m z) + \frac{2\mu}{m} \tilde{s}_{\mu\nu} \d_z J_\mu(m z) +  \frac{4}{k^2} k_{[\mu} \tilde{s}_{\nu]\rho} k^\rho \d_z [z J_{\mu-1}(m z)] \nonumber \\
=&~  \frac{2\mu}{m} \tilde{s}_{\mu\nu} \d_z J_\mu(m z) + 4 k_{[\mu}
\tilde{s}_{\nu]\rho} k^\rho \left[ \frac{z}{m} J_{\mu}(m z) + \frac{1}{k^2}
\d_z [z J_{\mu-1}(m z)]  \right]~,
\end{align}
while the components of $\frac{i \mu}{\ell} \star b$ are
\begin{align}\label{starbcomps}
\frac{i\mu}{\ell} (\star b)_{\mu\nu\rho} =&~ \frac{i\mu}{2 z} \cdot 2 \epsilon_{\mu\nu\rho}^{\phantom{\mu\nu\rho}\sigma} b_{\sigma z} \rightarrow - \frac{2 \mu}{m} \epsilon_{\mu\nu\rho}^{\phantom{\mu\nu\rho}\sigma} \tilde{s}_{\sigma \alpha} k^\alpha J_{\mu}(m z) \nonumber \\
=&~  -\frac{i \mu}{m} k_{\alpha} \epsilon_{\mu\nu\rho\sigma} \epsilon^{\sigma \alpha \beta\gamma} \tilde{s}_{\beta\gamma} J_{\mu}(m z)  = - \frac{2 i \mu}{m} (k_{\mu} \tilde{s}_{\nu\rho} + k_{\nu} \tilde{s}_{\rho \mu} + k_\rho \tilde{s}_{\mu\nu} ) J_{\mu}(m z)~,  \\
\frac{i\mu}{\ell} (\star b)_{\mu\nu z} =&~ - \frac{i \mu}{2 z} \epsilon_{\mu\nu}^{\phantom{\mu\nu}\rho\sigma} b_{\rho\sigma} \rightarrow - \frac{2 \mu^2}{m z} \tilde{s}_{\mu\nu} J_{\mu}(m z) - \frac{2 i \mu}{k^2} \epsilon_{\mu\nu}^{\phantom{\mu\nu}\rho\sigma} k_{\rho} \tilde{s}_{\sigma\alpha} k^\alpha J_{\mu-1}(m z)  \nonumber \\
=&~  - \frac{2 \mu^2}{m z} \tilde{s}_{\mu\nu} J_{\mu}(m z) + \frac{\mu}{k^2} k^\rho k_\alpha \epsilon_{\mu\nu\rho\sigma} \epsilon^{\sigma\alpha\beta\gamma} \tilde{s}_{\beta\gamma} J_{\mu-1}(m z) \nonumber  \\
=&~ \left[ - \frac{2\mu^2}{m z} J_{\mu}(m z) + 2\mu J_{\mu-1}(m z)
\right] \tilde{s}_{\mu\nu} + \frac{4\mu}{k^2} k_{[\mu} \tilde{s}_{\nu]\rho} k^\rho
J_{\mu-1}(m z)~.
\end{align}
Therefore we see that
\begin{equation}\label{munurhozero}
\left( db - \frac{i\mu}{\ell} \star b \right)_{\mu\nu\rho} = 0~,
\end{equation}
while
\begin{align}\label{munuznotzero}
\left( db - \frac{i\mu}{\ell} \star b \right)_{\mu\nu z} =&~ 2\mu \left[ \frac{1}{m} \d_z J_{\mu}(m z) + \frac{\mu}{m z} J_{\mu}(m z) - J_{\mu-1}(m z) \right] \tilde{s}_{\mu\nu} + \nonumber \\
& \qquad + \left[  \frac{k^2 z}{m} J_\mu(m z) + \d_z [ z J_{\mu-1}(m z) ] -\mu J_{\mu-1}(m z) \right] \frac{4}{k^2} k_{[\mu} \tilde{s}_{\nu] \rho} k^\rho  \nonumber \\
=&~ \frac{4 (k^2-m^2)}{m k^2}  k_{[\mu} \tilde{s}_{\nu] \rho} k^\rho z
J_{\mu}(m z)~,
\end{align}
after making use of Bessel function identities.

Then we have
\begin{align}\label{Lboneigen}
\bar{b} \wedge \left( db - \frac{i\mu}{\ell} \star b \right) =&~ \frac{1}{2! 3!} \bar{b}_{MN} \left( db - \frac{i\mu}{\ell} \star b \right)_{PQR} e^{MNPQR} d^5 x \nonumber \\
=&~ \frac{1}{4} \bar{b}_{\mu\nu} \left( db - \frac{i\mu}{\ell} \star b \right)_{\rho\sigma z} \epsilon^{\mu\nu\rho\sigma} d^5 x  \nonumber \\
=&~  \frac{1}{4} \cdot \frac{2\mu}{m'} \ \overline{\tilde{s}'}_{\mu\nu}(k) J_{\mu}(m' z) \left[ \frac{4 (k^2-m^2)}{m k^2}  k_{[\rho} \tilde{s}_{\sigma] \alpha} k^\alpha z J_{\mu}(m z) \right] \epsilon^{\mu\nu\rho\sigma} d^5 x \nonumber \\
=&~ - \frac{4 i \mu}{m m'} \left( k^\mu \overline{\tilde{s}'}_{\mu\rho} \right)
\frac{ (k^2 -m^2 )}{k^2} \left( k_\nu \tilde{s}^{\nu\rho} \right) z
J_{\mu}(m' z) J_{\mu}(m z) d^5 x~.
\end{align}
In the third step we plugged in some other, primed eigenmode for
$\bar{b}$ and noted that the longitudinal part of $\bar{b}$ does not
contribute to the contraction.  We must multiply this result by $-i$
when plugging into the action, and therefore the conjugate term gives a symmetric contribution, interchanging primed and unprimed modes.  The IR boundary condition leads to an orthogonal spectrum,
\begin{equation}\label{orthogonalmodes}
\int_{0}^{z_0} dz z J_{\mu}(m_n z) J_{\mu}(m_{n'} z) = \frac{z_{0}^2}{2} J_{\mu}(x_{\mu-1,n})^2 \delta_{n n'}~,
\end{equation}
and therefore, defining canonically normalized polarizations
\begin{equation}\label{canonicals}
s_{\mu\nu}^{(n)}(k) = \frac{z_0 J_{\mu}(x_{\mu-1,n}) }{g_b m_n} \sqrt{ \frac{2 \mu}{\ell} } \ \tilde{s}_{\mu\nu}^{(n)}(k)~,
\end{equation}
the free action in the $b$-sector takes the form
\begin{equation}\label{beigenaction}
S_{\rm sd} = - \sum_{n} \int
\frac{d^4k}{(2\pi)^4} \tr \left\{  \left( k^\mu \bar{s}_{\mu\rho}^{(n)} \right)
\frac{ (k^2 -m_{n}^2 )}{k^2} \left( k_\nu s^{(n)\nu\rho} \right) \right\}~.
\end{equation}
Note that, since $[b] = 3/2$, we have $[\tilde{s}] = 5/2$ and therefore $[s] = 1$.

It is natural that only the longitudinal components of $s$ should enter into \eqref{beigenaction}: $s$ is a self-dual tensor, so the transverse components are not independent degrees of freedom.  The six real degrees of freedom contained in $s$ may be taken, in the rest frame, as ${\rm Re}(s_{0i})$ and ${\rm Im}(s_{0i})$.  According to our charge and parity assignments for $b$, discussed in section 4, the imaginary-longitudinal (equivalently, real-transverse) components of $s$ represent $h_1/b_1$ modes, while the real-longiudinal (equivalently, imaginary-transverse) components of $s$ represent tensor $\omega/\rho$ modes, \eqref{Oeven}.  Such modes can be given a canonical quadratic action by embedding them into $s$ according to
\begin{equation}\label{sembed}
s_{\mu\nu}^{(n)}(k) = P_{\mu\nu}^{+ \alpha\beta} \left[ - \frac{2i}{k} \epsilon_{\alpha\beta}^{\phantom{\alpha\beta}\delta\gamma} k_\delta \left( b_{1\gamma}^{(n)}(k) + i \rho^{T,(n)}_{\gamma}(k) \right) \right]~.
\end{equation}
As we discussed, at the level of the free action, these modes have identical spectra.


\end{document}